\documentclass[fleqn,usenatbib]{mnras}

\usepackage{newtxtext,newtxmath}
\usepackage[T1]{fontenc}
\usepackage[normalem]{ulem} %% to use \sout
\usepackage{xcolor}

\DeclareRobustCommand{\VAN}[3]{#2}
\let\VANthebibliography\thebibliography
\def\thebibliography{\DeclareRobustCommand{\VAN}[3]{##3}\VANthebibliography}

\usepackage{graphicx}	% Including figure files
\usepackage{amsmath}	% Advanced maths commands

\usepackage{amssymb}	% Extra maths symbols

\title[Maximum temperatures and dust compositions]{Maximum Temperatures in Evolving Protoplanetary Discs and Composition of Planetary Building Blocks}

\author[M. Li et al.]{
Min Li,$^{1}$
Shichun Huang,$^{2}$
Zhaohuan Zhu,$^{1}$
Michail I. Petaev,$^{3,4}$
and Jason H. Steffen$^{1}$\thanks{E-mail: \href{mailto:minphyli@163.com}{minphyli@163.com} (ML); \href{mailto:jason.steffen@unlv.edu}{jason.steffen@unlv.edu} (JHS)}
\\
$^{1}$Department of Physics and Astronomy, University of Nevada, Las Vegas, 4505 S. Maryland Pkwy, Las Vegas 89154, USA\\
$^{2}$Department of Geoscience, University of Nevada, Las Vegas, 4505 S. Maryland Pkwy, Las Vegas 89154, USA\\
$^{3}$Department of Earth \& Planetary Sciences, Harvard University, 20 Oxford St., Cambridge 02138, USA\\
$^{4}$Harvard-Smithsonian Center for Astrophysics, 60 Garden St., Cambridge 02138, USA
}

\date{Accepted XXX. Received YYY; in original form ZZZ}

\pubyear{2020}

\begin{document}
\label{firstpage}
\pagerange{\pageref{firstpage}--\pageref{lastpage}}
\maketitle

\begin{abstract}
The maximum temperature and radial temperature profile in a protoplanetary disc are important for the condensation of different elements in the disc.  We simulate the evolution of a set of protoplanetary discs from the collapse of their progenitor molecular cloud cores as well as the dust decoupling within the discs as they evolve.  We show how the initial properties of the cloud cores affect the thermal history of the protoplanetary discs using a simple viscous disc model.  Our results show that the maximum midplane temperature in the disc occurs within 0.5 AU.  It increases with the initial cloud temperature and decreases with its angular velocity and the viscosity of the disc.  From the observed properties of the molecular cloud cores we find the median value of the maximum temperature is around 1250 K, with roughly 90\% of them being less than 1500 K -- a value that is lower than the 50\% condensation temperatures of most refractory elements.  Therefore, only cloud cores with high initial temperatures or low angular velocities and/or low viscosities within the planet-forming discs will result in refractory-rich planetesimals.  To reproduce the volatile depletion pattern of CM, CO, and CV chondrites and the terrestrial planets in Solar system, one must either have rare properties of the initial molecular cloud cores like high core temperature, or other sources of energy to heat the disc to sufficiently high temperatures.  Alternatively, the volatile depletion observed in these chondrites may be inherited from the progenitor molecular cloud.
\end{abstract}

\begin{keywords}
stars: pre-main-sequence -- accretion, accretion discs -- astrochemistry -- planets and satellites: composition -- protoplanetary discs
\end{keywords}

%%%%%%%%%%%%%%%%%%%%%%%%%%%%%%%%%%%%%%%%%%%%%%%%%%

%%%%%%%%%%%%%%%%% BODY OF PAPER %%%%%%%%%%%%%%%%%%

\section{Introduction}

If planets and their central star form from the same nebular reservoir, their chemical compositions must correlate in some fashion.
However, observed planet/planetesimal compositions do not completely match those of their central stars.  For example, in our solar system, 
all rocky
planets and planetesimals contain near-solar proportions of
refractory elements but are depleted in volatile elements
\citep{Asplund:2005,Asplund:2009}. 
Such volatile depletion is thought
to result from planetesimal formation in the solar nebula
\citep[][and see \citet{Li:2020} for a summary]{Cassen:1994,Cassen:1996,Cassen:2001,Ciesla:2008,Bond:2010a,Bond:2010b,Elser:2012,Moriarty:2014,Pignatale:2016}  
where they form by coalescence of condensed mineral
dust
(See \citet{Johansen:2014} for a review).  The planetesimal composition at a given location depends upon the initial chemical composition and the evolution of the disc \citep{Cassen:1996,Ciesla:2008,Li:2020}.
The same is likely true for
exoplanetary systems.

As the initial, cold molecular cloud cores collapse, discs form due to the existing angular momentum of the system \citep[e.g.][]{Nakamoto:1994,Sui:2019}.  Subsequent heating from the forming central star and from the viscosity of the disc can vaporise primordial condensed material -- forcing the condensation sequence to begin anew, starting under high temperature and pressure conditions that gradually lessen as the disc evolves and dissipates.  However, some elements condense at relatively high temperatures \citep[e.g.][]{Li:2020}.  Depending upon the properties of the initial molecular cloud core (MCC) and the disc that forms therefrom, the temperatures that arise in the disc may not be sufficient to vaporise the most refractory of the pre-existing material.  This situation presents a problem when trying to understand the observed chemistry of planet-forming materials since certain minerals contained in planetesimals may form in different conditions, times, and places.

Using an analytical approach, \citet{Cassen:1996} showed that the depletion of moderate volatile elements observed in CM, CO and CV carbonaceous chondrites can be reproduced using a simple disc model.
Alternatively, \citet{Yin:2005} suggested that the observed depletion pattern was inherited from the molecular cloud in which the gas is depleted in refractory elements that have already condensed into grains.  Modeling the migration of solids in the viscously evolving discs, \citet{Ciesla:2008} found that the temperature should be higher than $\sim$ 1350 K at 2 to 4 AU to explain the
volatile depletion patterns in CM, CO and CV chondrites from the asteroid belt, which is difficult to achieve in a traditional alpha protoplanetary disc model.
In a subsequent paper, \citet{Yang:2012} investigated the mixing process of the grains in the solar nebula. They found that, during the evolution of solar nebula, the fraction of refractory elements is the largest when the infall from the molecular cloud core stops (at age of several $10^5$ years).

More recently, \citet{Li:2020} re-visited the \citet{Cassen:1996} results by combining the original disc P-T evolution with a chemical equilibrium model \citep[GRAINS,][]{Petaev:2009}.  They reproduced the enrichment in refractory and depletion in volatile elements observed in CM, CO and CV chondrites. However,
the temperature of the disc model used in their work and by \citet{Cassen:1996} may be unrealistically high (see also \citet{Ciesla:2008}).  Here we examine the temperature and pressure histories of the midplanes of protoplanetary discs that form from the collapse of molecular cloud cores as a function of the initial properties of cores to gain a better understanding of its potential effects on the dust condensation process. We select our initial conditions from the observed MCC properties to provide a realistic estimate for the properties of the resulting discs. Specifically, we consider the distribution of the initial 
temperature and angular velocity of the cores, and viscosity in the disc throughout their evolution.

Our initial cores self-consistently collapse to form stars with surrounding protoplanetary discs that evolve under the influence of stellar irradiation and
viscosity -- which may arise from gravitation instability (GI) at early stage of disc evolution \citep{Lin:1987,Lin:1990,Gammie:2001,Clarke:2009,Rice:2009,Rice:2010}, magnetorotational instability \citep[MRI;][]{Balbus:1991} and hydrodynamic processes \citep{Dubrulle:1993,Dubrulle:2005}.  We use a constant $\alpha$ for the viscosities in this paper.
We calculate the temperature and pressure conditions in these discs and compare them to the condensation temperatures of a variety of elements calculated by \citet{Li:2020}.  From this information we make statements regarding the origins of some compounds that are or could be observed in planetary systems, including the solar system.
Our constant alpha disc model provides a benchmark for comparing with previous studies \citep[e.g.][]{Cassen:1996,Ciesla:2008} and future studies. 
A more realistic, layered accretion model is left for elsewhere.

This paper is organized as follows.  We begin by outlining our parameters for the initial molecular cloud cores and the evolution model of the protoplanetary disc in Section \ref{sec:clouddisc}.  We show the evolution of the disc in Subsection \ref{sec:discevo}.  The dependence of the maximum temperature in discs on the properties of MCC and the statistical properties of the maximum temperatures are shown in \ref{sec:maxtem}.  We discuss the implications of our results on the elemental composition of chondrites and the formation of species in Section \ref{sec:discussion} and give our concluding remarks in Section \ref{sec:conclusion}.

\section{Molecular Cloud Core and Disc Evolution} \label{sec:clouddisc}

\begin{figure}
	\includegraphics[width=\columnwidth]{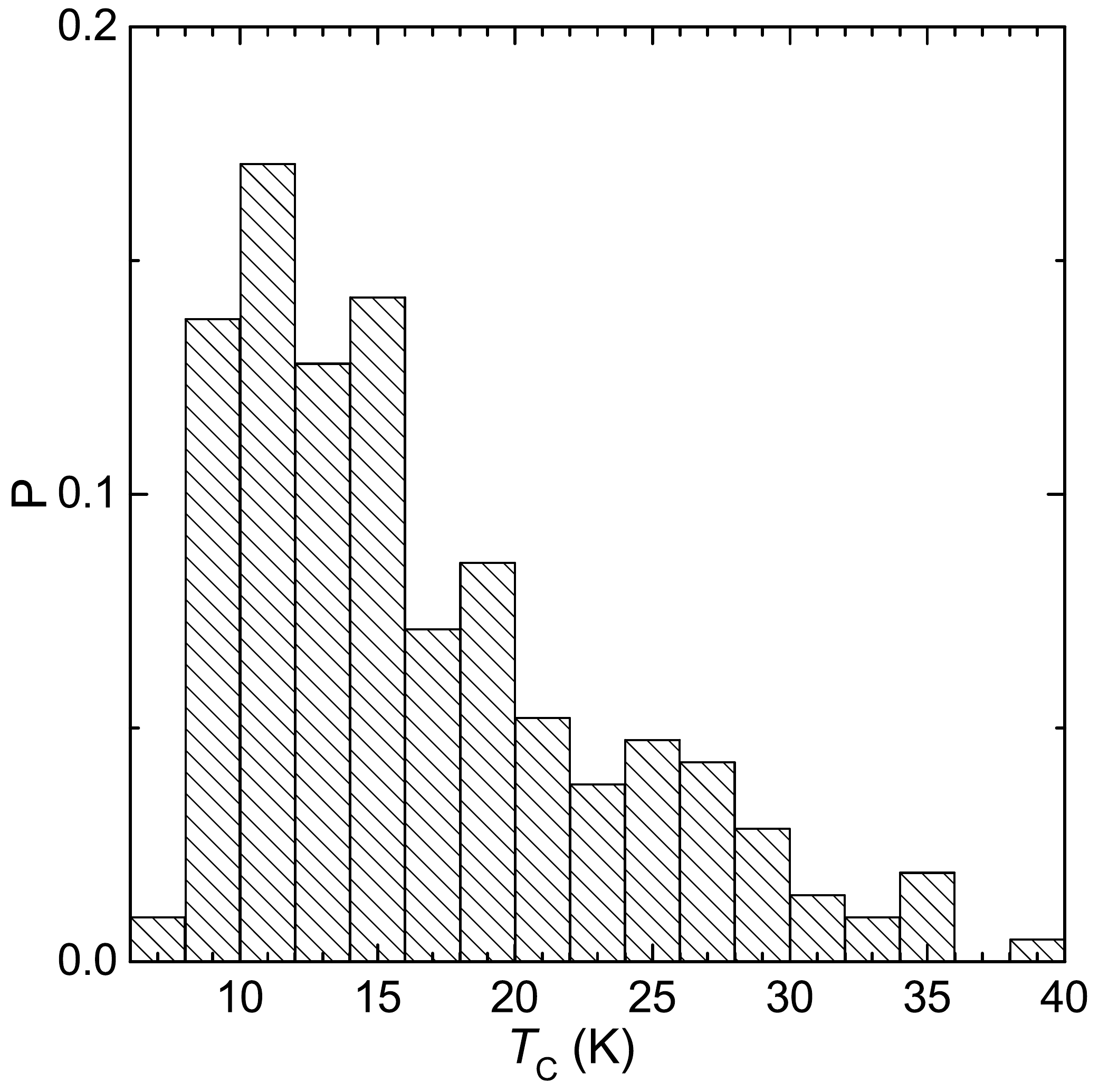}
	\includegraphics[width=\columnwidth]{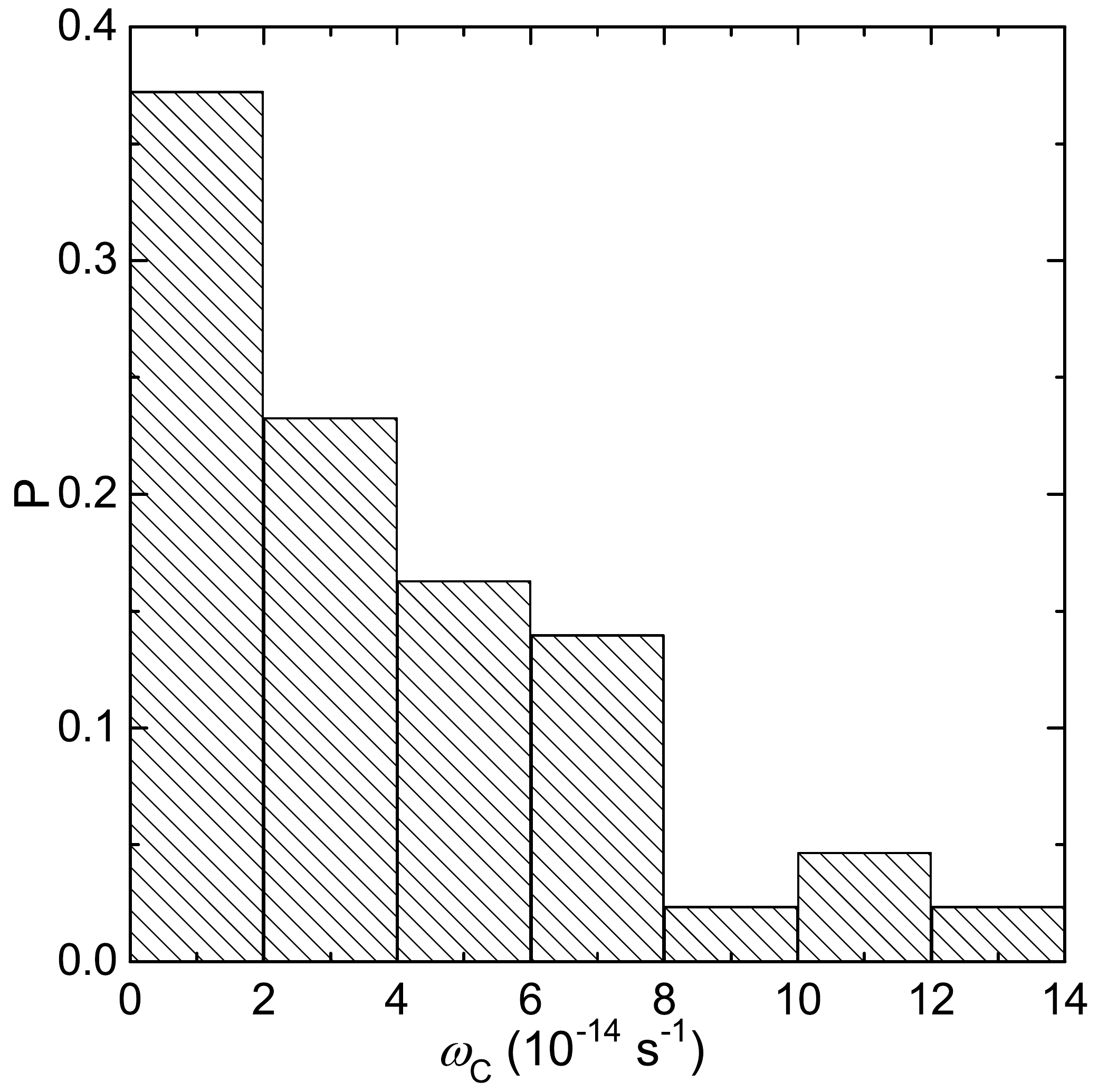}
    \caption{Distribution of $T_{\rm C}$ (upper panel) and $\omega_{\rm C}$ (lower panel) of observed molecular cloud cores. This results are from 211 temperature values \citep{Jijina:1999} and 43 angular velocities \citep{Goodman:1993}.}
    \label{fig:tem}
\end{figure}

Our simulations begin with the collapse of a molecular cloud core, which self-consistently evolves into a protoplanetary disc.  For all of these simulations we consider cores of one solar mass.  The initial temperature ($T_{\rm C}$) and angular velocity ($\omega_{\rm C}$) of each core is chosen from the observed distributions of these properties shown in Figure \ref{fig:tem} \citep[see also  Figures 1 and 2 of][]{Li:2015}.  The temperatures of the cores range from roughly 7 to 40 Kelvins with the median value of 15 K \citep{Jijina:1999}. The angular velocities of the cores range from 0.3 to 13 $\times \ 10^{-14}\rm\ s^{-1}$ \citep{Goodman:1993}.  Since very little mass either escapes the system or is incorporated into planets, we assume that the final mass of the central star is equal the mass of the MCC.

The evolution of the protoplanetary disc that forms from the collapse of an MCC is governed by \citep{Li:2015}
\begin{equation}\label{equ.diff}
\begin{aligned}
\frac{\partial \Sigma(R,t)}{\partial t}
=&\frac{3}{R} \frac{\partial}{\partial R} \left[ R^{1/2} \frac{\partial}{\partial R} (\Sigma \nu R^{1/2}) \right] +S(R,t)\\
&+S(R,t)\left\{2-3\left[\frac{R}{R_{d}(t)}\right]^{1/2}+\frac{R/R_{d}(t)}{1+[R/R_{d}(t)]^{1/2}}\right\}.
\end{aligned}
\end{equation}
Here $\Sigma(R,t)$ is the gas surface density of the disc at radius $R$ and time $t$ while $\nu$ is the kinematic viscosity. 
The third term on the right hand side of Equation (\ref{equ.diff}) is due to the difference of the specific angular momentum between the infalling material and the material in the disc.

The mass influx onto the disc and protostellar system, $S(R,t)$, is \citep{Nakamoto:1994}:
\begin{equation}\label{equ.inf}
S\left(R,t\right)=
\begin{cases}
{\displaystyle \frac{\dot{M}}{4\pi R R_{\mathrm{d}}\left(t\right)}
\left[1-\frac{R}{R_{\mathrm{d}}\left(t\right)}\right]^{-1/2} } &{\rm if\ } {\displaystyle \frac{R}{R_{\mathrm{d}}\left(t\right)}<1; } \\
{}\\
0 &{\rm otherwise},
\end{cases}
\end{equation}
where $\dot{M}$ is the mass infall rate \citep{Shu:1977}
\begin{equation}\label{equ.mdot}
\dot{M} = \frac{0.975}{G}\left(\frac{\mathcal{R}}{\mu}\right)^{3/2}T_{\rm C}^{3/2},
\end{equation}
where $G$ is the gravitational constant, $\mathcal{R}$ is the gas constant, $\mu = 2.33$ is the mean molecular mass.  Time $t=0$ is set to be the time when MCC starts to collapse.

The centrifugal radius is
\begin{equation}\label{equ.rd}
R_{\mathrm{d}}(t) = 31
\left(\frac{\omega_{\rm C}}{10^{-14} {\rm\ s} ^{-1} } \right)^{2}
\left(\frac{T_{\rm C}}{10 {\rm\ K}} \right)^{1/2}
\left(\frac{t}{5\times10^{5} {\rm\ yr}} \right)^{3} {\rm AU},
\end{equation}
where $\omega_{\rm C}$ is the angular momentum of the MCC.

We use the $\alpha$-prescription \citep{Shakura:1973} to calculate the viscosity, $\nu=\alpha c_s H$, where $\alpha$ is a dimensionless parameter less than 1, $H$ is the half thickness of the gas disc, $c_s=\sqrt{\mathcal{R} T/\mu}$ is the sound speed, and $T$ is the temperature of the mid-plane of the disc. 
Observations indicate that the $\alpha$ value has a large range and there is no preferred value of it \citep{Rafikov:2017}.
For our purposes, each simulation uses a constant $\alpha$, and the suite of simulations examines values equal to $10^{-1}$, $10^{-2}$, $10^{-3}$, $10^{-4}$, and $10^{-5}$.

The disc surface temperature is determined by the thermal equilibrium between the cooling and heating fluxes of the disc, which is \citep{Hueso:2005}
\begin{equation}\label{equ.ts}
\sigma T_{s}^{4}=\frac{1}{2}
\left(1+\frac{1}{2\tau_{P}}\right)(\dot{E}_{\nu}+\dot{E}_{s})+\sigma
T_{\rm ir}^{4} +\sigma T_{\rm C}^{4},
\end{equation}
where $\sigma$ is the Stefan–Boltzmann constant, $\tau_{P}=\kappa_{P}\Sigma$ is the Planck mean optical depth, where $\kappa_{P}$ is the Planck mean opacity. $\dot{E}_{\nu}$ is the viscous dissipation rate and $\dot{E}_{s}$ is the energy generation rate by shock heating 
-- the energy difference between the cloud core and the disc when the infalling material merges with the disc.
\citep{Nakamoto:1994}.  $T_{\rm ir}$ is the effective temperature due to the irradiation from the protostar.

The midplane temperature in the disc is
\begin{equation}\label{equ.tm}
\sigma T^{4}=\frac{1}{2}
\left[\left(\frac{3}{8}\tau_{R}+\frac{1}{2\tau_{P}}\right)\dot{E}_{\nu}
+\left(1+\frac{1}{2\tau_{P}}\right)\dot{E}_{s}\right] 
+\sigma T_{\rm ir}^{4} +\sigma T_{\rm C}^{4},
\end{equation}
where $\tau_{R}=\kappa_{R}\Sigma$ is the Rosseland mean optical depth, and $\kappa_{R}$ is the Rosseland mean opacity. Here $\kappa_{P}=2.39\kappa_{R}$ \citep{Nakamoto:1994}.
We use the same method as in \citet{Armitage:2001} to calculate the opacity, which comes from \citet{Bell:1994} for high temperature and \citet{Bell:1997} for low temperature.  
The inner radius of the subsequent disc is 0.3 AU. At the inner radius, we set $\Sigma$ to be 0, and the material is accreted to the central star \citep{Bath:1981,Lin:1987}. The outer radius is 1.25e5 AU -- which allows the disc to expand freely.

\section{Results}
\label{sec:results}

\subsection{Disc evolution}
\label{sec:discevo}

\begin{figure}
	\centering  
	\includegraphics[width=0.8\columnwidth]{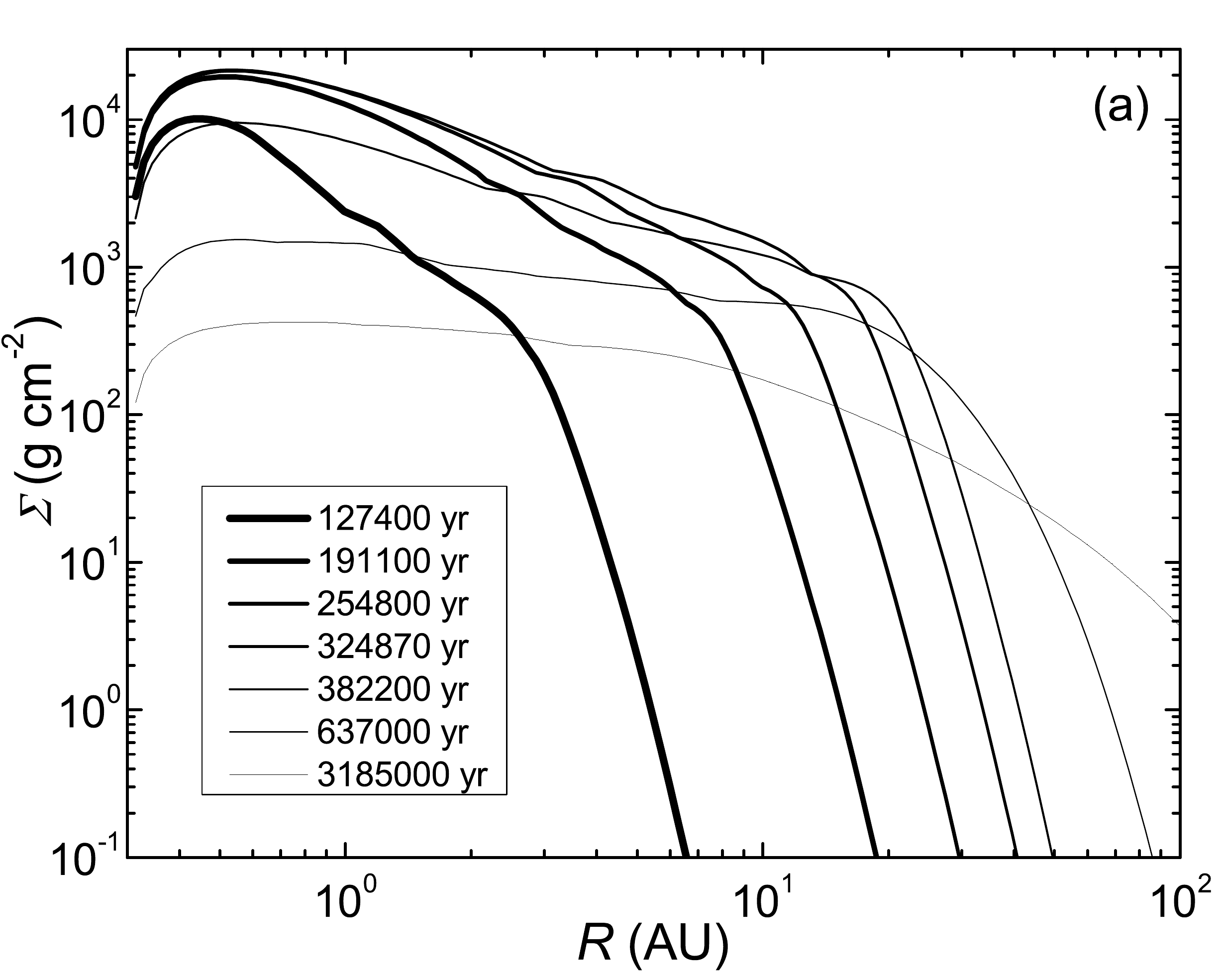}
	\includegraphics[width=0.8\columnwidth]{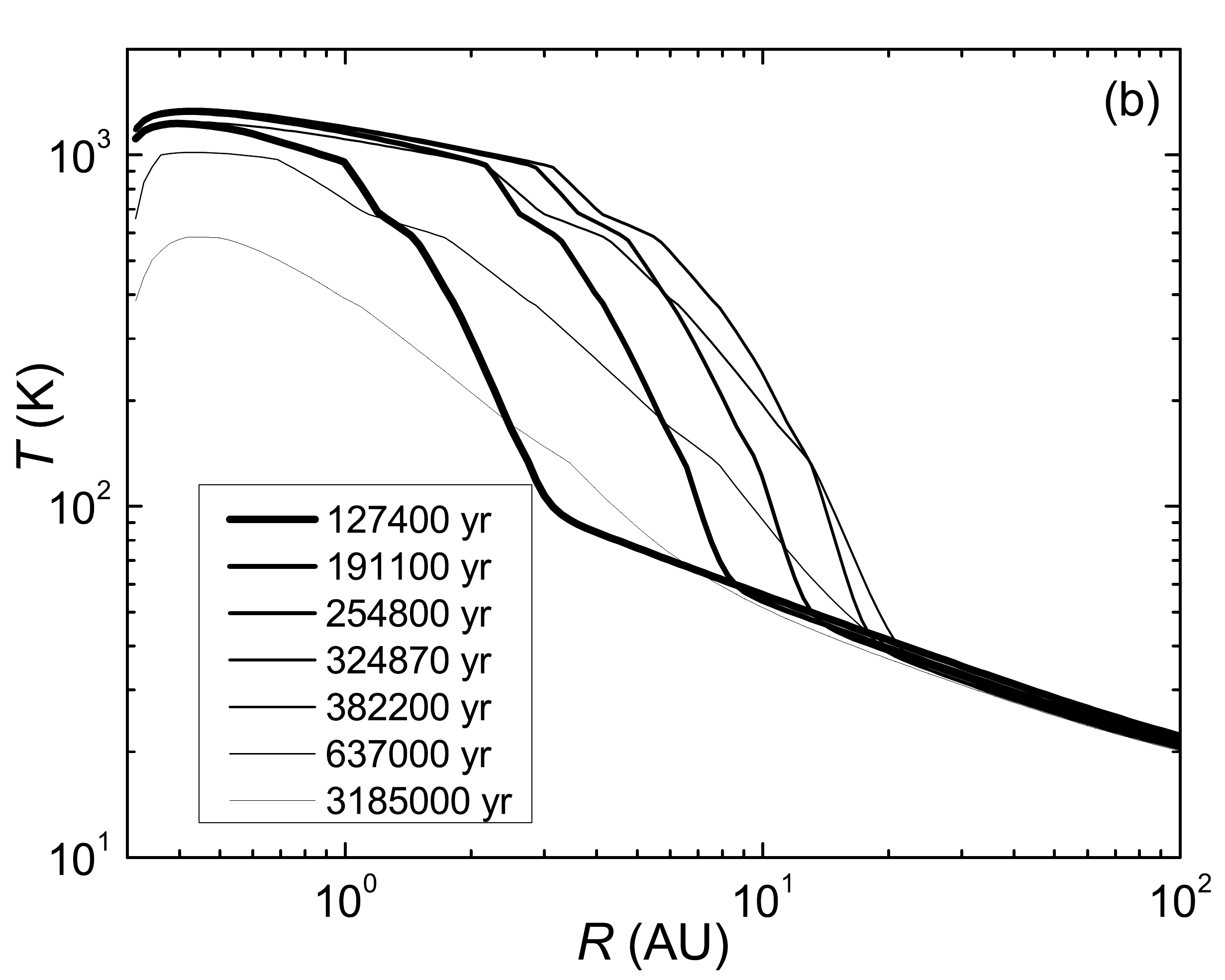}
	\includegraphics[width=1.15\columnwidth]{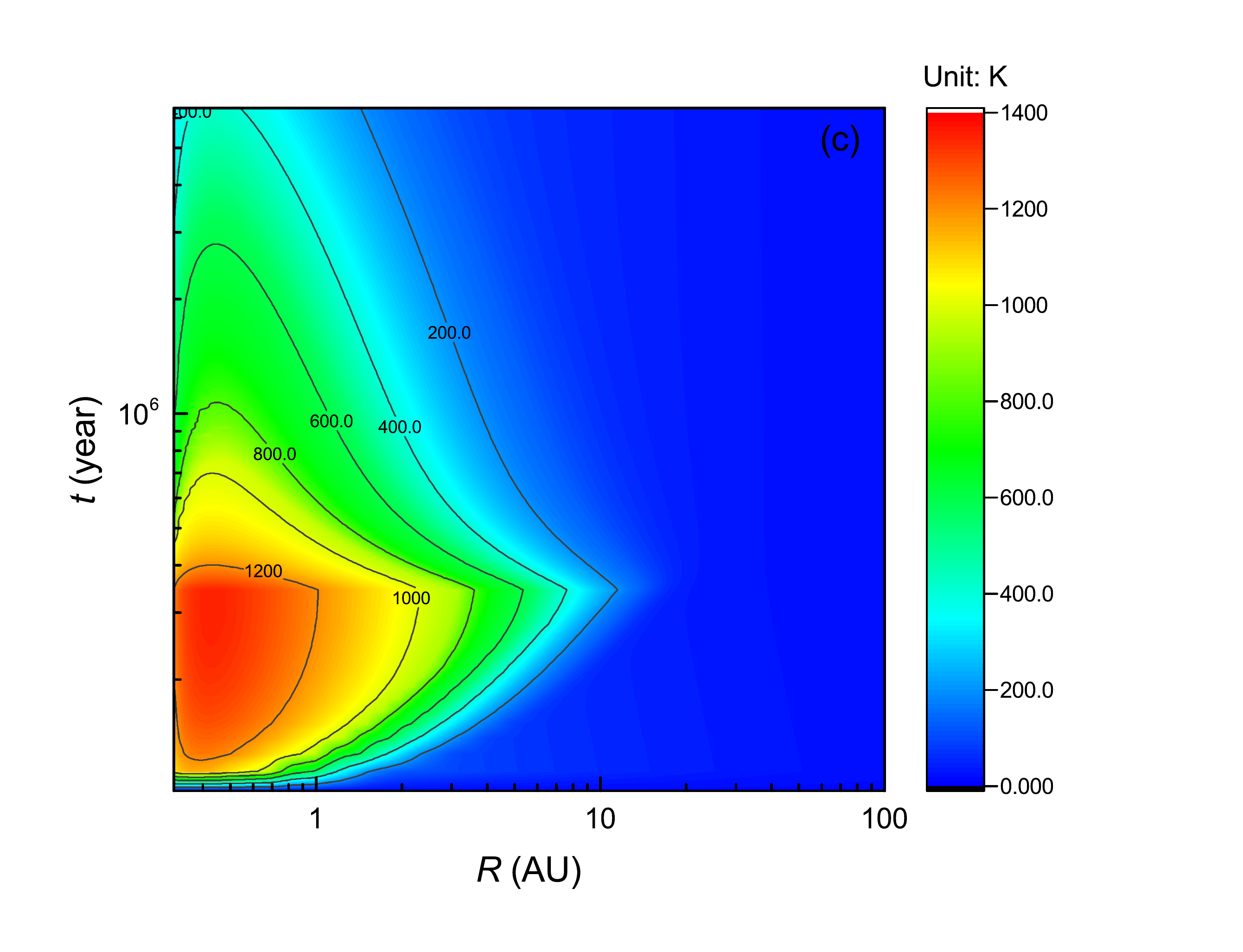}
    \caption{Evolution of (a) surface density and (b) midplane temperature as a function of radius and time.  (c)  Contour plot of temperature as a function of radius and time.  Here, $M_{\rm C}=1\ M_\odot$, $T_{\rm C}=15$ K, $\omega_{\rm C}=1\times 10^{-14} \rm\ s^{-1}$, and $\alpha=10^{-3}$.  Note that the temperature gets its maximum value, 1350 K, at $t=$ 324870 yr at 0.43 AU.}
    \label{fig:discevo}
\end{figure}

The initial conditions of the MCC and the chosen $\alpha$ value determine the evolution of the disc surface density, midplane temperature, and the disc evolution timescale.  Much of the discussion that follows uses a fiducial model for the properties of the MCC with a mass $M_\text{C} = 1\ M_\odot$, initial temperature of $T_\text{C} = 15$ K, an initial angular velocity of $\omega_\text{C} = 1 \times 10^{-14} \ \text{s}^{-1}$, and $\alpha=1\times 10^{-3}$. 

Figure \ref{fig:discevo} shows how the disc surface density and midplane temperature evolve with time. At very early times, the surface density (Figure \ref{fig:discevo}a) increases at all radii as material is supplied by the collapsing MCC.  It reaches the maximum value in the inner region (within 10 AU) at around the time when the collapse ceases (3.47$\times 10^5$ years).  After that, it decreases quickly in the inner region and increases in the outer region as the disc expands from the effects of its viscosity.  

The general trends of the temperature evolution (Figure \ref{fig:discevo}b) are similar to that of the surface density except that the temperatures at radii larger than 20 AU do not change much with time.  Temperatures at larger radii peak at a later time than those at smaller radii (see the direction of contour in Figure \ref{fig:discevo}c).  For the fiducial case, the temperature in the disc reaches its maximum value of 1350 K at 0.43 AU and $t=$ 324870 yr (which is $0.937 \simeq 1$ times the collapse timescale of the MCC).

\subsection{Maximum disc temperature as a function of molecular cloud core properties}
\label{sec:maxtem}

We first consider the maximum temperatures in the disc as a function of the initial MCC properties -- specifically temperature and angular velocity for a solar mass core.  As the properties of the MCC change, the mass infall rate and the centrifugal radius change (See Equations \ref{equ.inf} and \ref{equ.rd}).  Therefore, the evolution of the disc and the maximum temperature in the disc also change.  For our study, we examine initial temperatures ranging from 7 to 39 K, and initial angular velocities ranging from 1 to 13 $\times 10^{-14}$ s$^{-1}$, and $\alpha$ values ranging from $10^{-5}$ to $10^{-1}$
\citep{Rafikov:2017}.

\begin{figure}
	\includegraphics[width=\columnwidth]{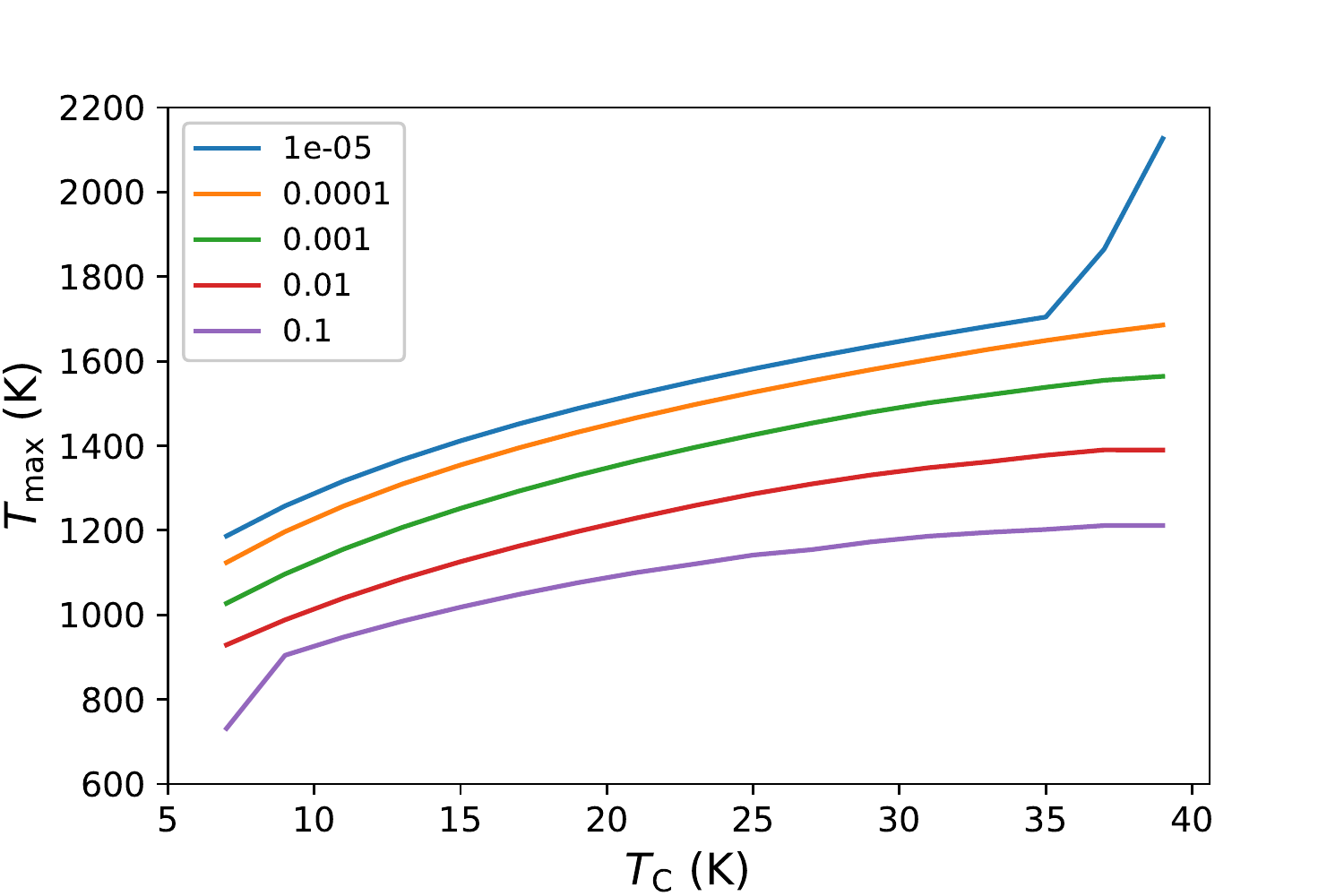}
    \caption{Maximum temperature as a function of  $T_{\rm C}$ for different $\alpha$. Here $M_{\rm C}=1\ M_\odot$ and $\omega_{\rm C}=3\times 10^{-14} \rm s^{-1}$.}
    \label{fig:T-t}
\end{figure}

Figure \ref{fig:T-t} shows the maximum midplane temperature reached by a disc as a function of $T_{\rm C}$ for different $\alpha$ where the initial MCC angular velocity is fixed at $\omega_{\rm C}=3\times 10^{-14} \rm s^{-1}$.  The maximum temperature increases with $T_{\rm C}$, but decreases with the viscosity.  For each $\alpha$, as $T_{\rm C}$ increases, the mass infall rate increases (Equation \ref{equ.inf}), and there will be more material at the inner region of the disc.  Thus, the surface density in the inner region and the maximum temperature increase.  As $\alpha$ decreases, the efficiency of the expansion of the disc decreases, leaving more materials in the inner region of the disc and also a higher maximum temperature.  However, the dependence of the maximum temperature on viscosity is relatively weak -- changing by only $\sim$ 50\% over four orders of magnitude in $\alpha$.  (Though the resulting differences happen to span an important range near the condensation temperatures of many elements.)  For the highest temperature MCCs, we see a significant increase in the highest disc temperatures
as there is more material in the inner regions of the disc -- raising the opacity and keeping the heat within the disc.
For all the cases, the maximum temperatures are between $\sim$750 and 2200 K.

\begin{figure}
	\includegraphics[width=\columnwidth]{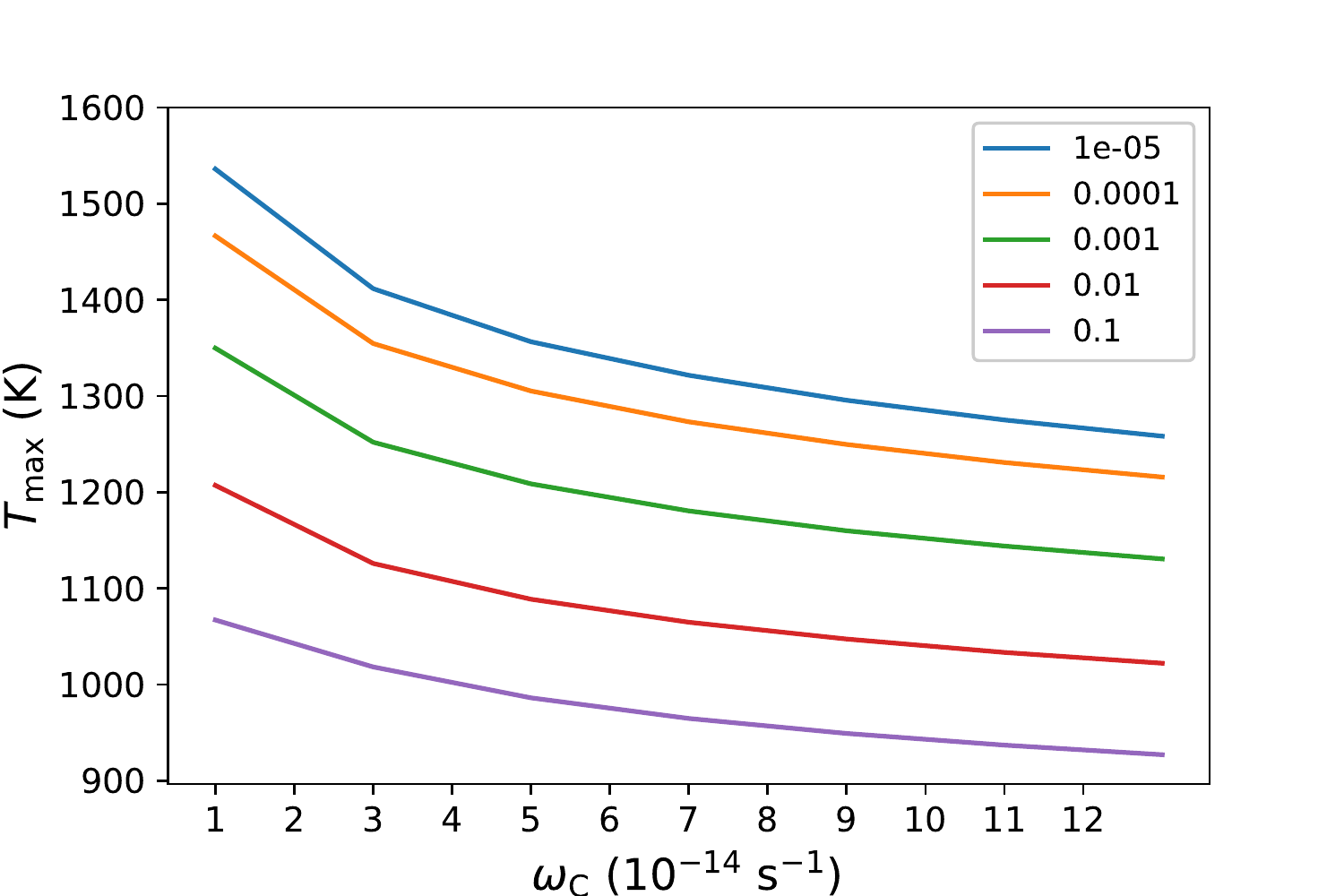}
    \caption{Maximum temperature as a function of  $\omega_{\rm C}$ for different $\alpha$. Here, $M_{\rm C}=1\ M_\odot$, $T_{\rm C}=15$ K.}
    \label{fig:T-w}
\end{figure}

Figure \ref{fig:T-w} shows the maximum disc temperature for an initial MCC temperature of 15 K, as a function of the initial MCC angular velocity -- again for a variety of $\alpha$.  For each $\alpha$, as $\omega_{\rm C}$ increases, the centrifugal radius increases (Equation \ref{equ.rd}), there will be less material in the inner regions relative to the outer regions, and the maximum midplane temperature decreases.  As with Figure \ref{fig:T-t}, the disc with lower $\alpha$ has higher maximum temperature, and the max temperatures range from 900 to 1600 K.

\subsection{Statistical properties of maximum disc temperatures}

\begin{figure}
    \includegraphics[width=\columnwidth]{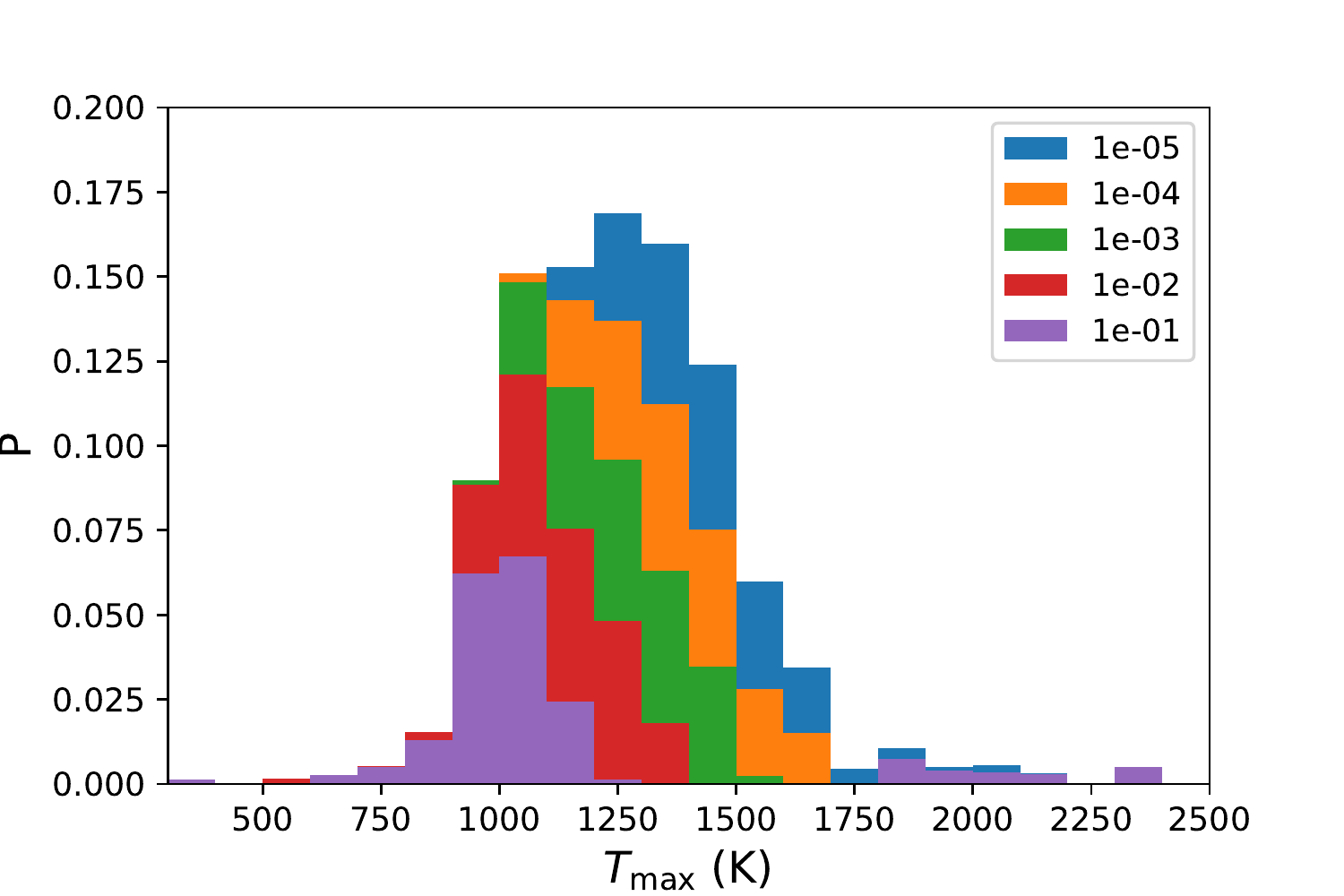}
    \caption{Stacked histogram of maximum temperatures achieved in our ensemble of discs.
    Different color means contribution from different $\alpha$.}
    \label{fig:N_tmax}
\end{figure}

We now examine the maximum midplane temperatures that would arise in planet-forming discs, given initial conditions that match the observed properties of MCCs.  For temperature $T_{\rm C}$, we choose values from 7 to 39 K with intervals of 2 K.  For $\omega_{\rm C}$, the values are from 1 to 13 $\times\ 10^{-14} \rm\  s^{-1}$ with an interval of 2 $\times\  10^{-14} \rm\ s^{-1}$ \citep{Li:2015}.
For $\alpha$, we adopt $10^{-1}$, $10^{-2}$, $10^{-3}$, $10^{-4}$, or $10^{-5}$, distributed equally among the 595 simulations \citep{Rafikov:2017}.  In all of our simulations, the maximum midplane temperatures are reached at radii less than 1 AU and at early times following the collapse of the MCC.  Given the distributions of $T_{\rm C}$, $\omega_{\rm C}$, and $\alpha$, the distribution of maximum temperatures in the disc, weighted by these initial conditions (Figure \ref{fig:tem}), is shown in Figure \ref{fig:N_tmax}.

For the majority of discs in the regime of these parameters, the maximum temperature is between 1000 and 1500K, with the median value being around 1250 K.  This value of the maximum midplane temperature is lower than the 50\% condensation temperature of silicon at a total pressure of $10^{-4}$ bars
\citep[$\sim$ 1300 K,][]{Lodders:2003,Li:2020}.
This result has important implications for the evolution
of primordial MCC dust because all refractory elements with higher condensation temperatures will not evaporate and fractionate from each other.  Meanwhile, the moderately and highly volatile elements would be evaporated and fractionated from the condensed elements if gaseous and condensed phases are physically separated.

The resulting heterotemporal condensation sequence, barring an additional heating mechanism, means that all refractory elements and most moderately volatile elements would be locked into dust from the beginning.  
Only highly volatile elements would be affected at later times, during the evolution of the disc.  In addition, the low maximum midplane temperatures inferred from our model may not even allow refractory inclusions, including CAIs, to form inside the evolving protoplanetary disc, since many refractory inclusion require much higher temperatures (>2000 K) to form fractionated rare-earth-element patterns \citep[e.g.,][]{MacPherson:2003,Petaev:2009b}.

\begin{figure}
	\includegraphics[width=\columnwidth]{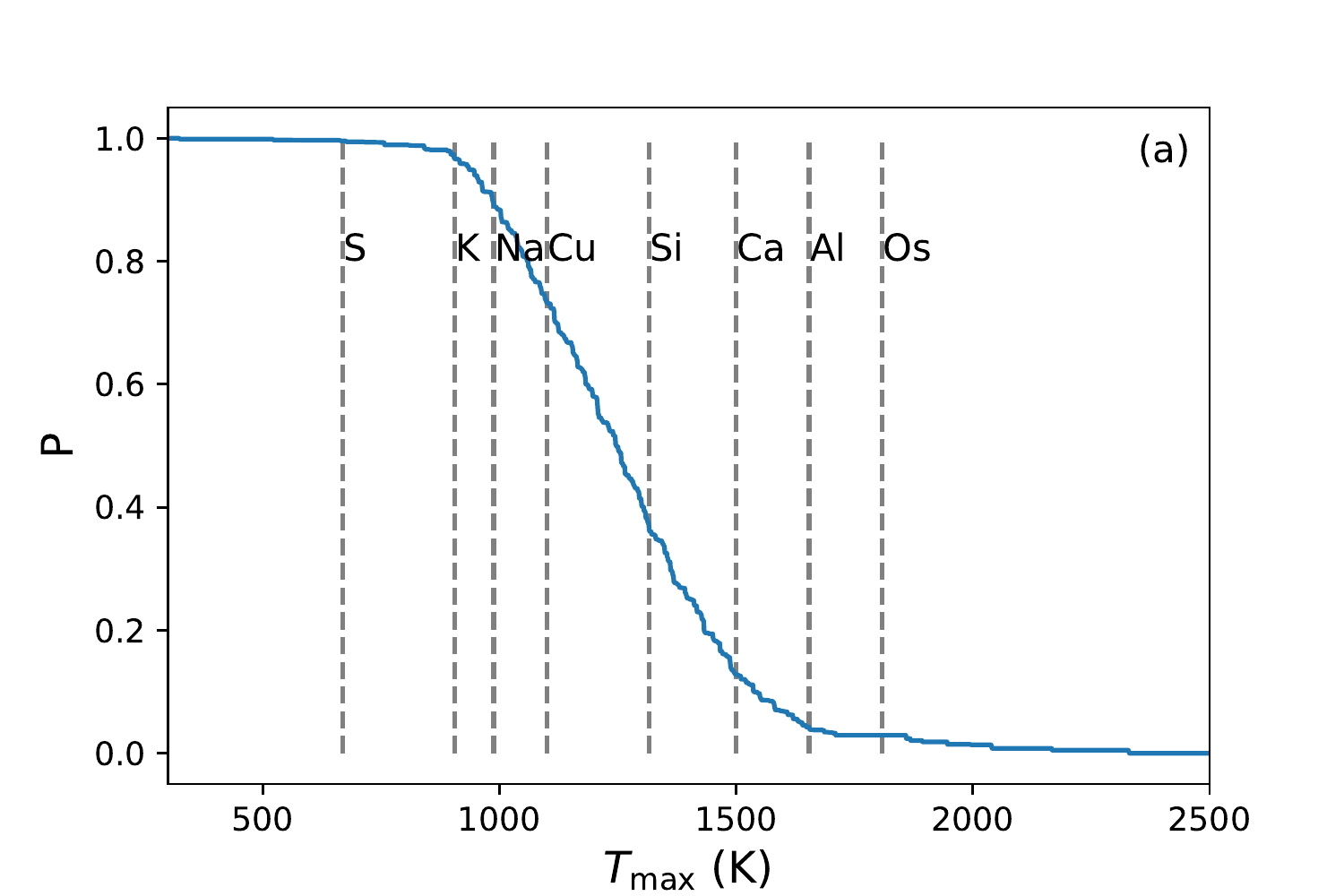}
	\includegraphics[width=1.0\columnwidth]{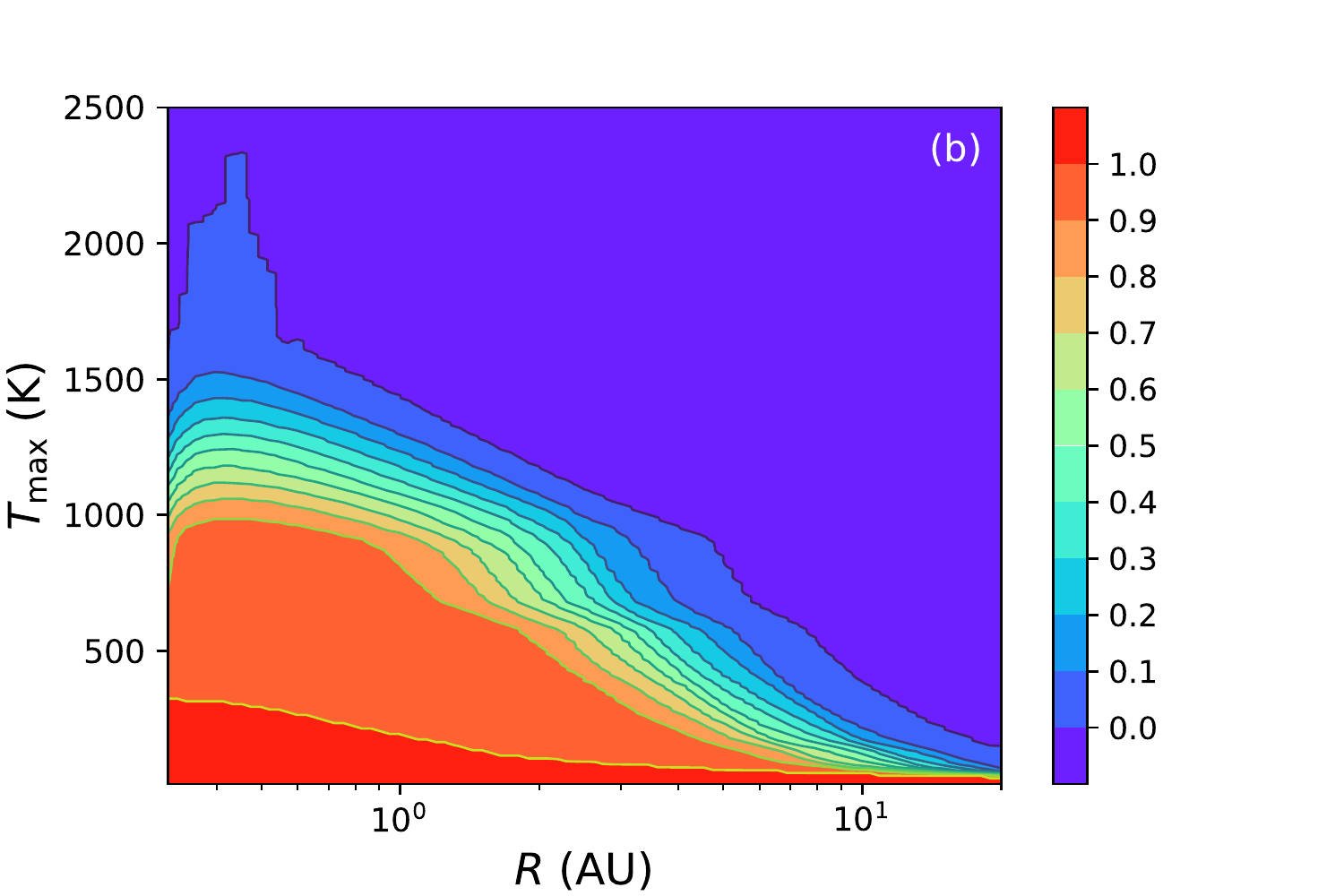}
    \caption{(a) Fraction of discs that reach a maximum temperature higher than a given value.  The vertical lines are the 50\% condensation temperatures at $10^{-4}$ bar for different elements \citep{Li:2020}.
    (b) Fraction of discs that reach specific peak temperatures as a function of radius.}
    \label{fig:p_tmax}
\end{figure}

Another interesting quantity to consider is the fraction of systems that would achieve a maximum temperature above a given temperature, beginning with the observed MCC distributions.  In Figure \ref{fig:p_tmax}, we plot the accumulated probability of the maximum temperatures in our discs  
as a function of radius.  Figure \ref{fig:p_tmax}a shows that 5\% of the discs reach maximum temperatures lower than 935 K and another 5\% reach maximum temperatures higher than 1635 K (near the 50\% condensation temperature of Al at $10^{-4}$ bar). 
Note that different $\alpha$ results in different maximum temperatures,
and therefore the distribution of $\alpha$ will affect the distribution of maximum temperatures.

Few discs in our sample reach maximum temperatures higher than 2000 K.  These temperatures only occur with the highest temperature and lowest viscosity MCCs.  We also plot the 50\% condensation temperature at $10^{-4}$ bar for several elements for comparison purposes.  From these results, we see that the maximum temperatures for most discs are less than the 50\% condensation temperature of the most refractory elements, such as Ca, Al, and Os.  Moreover, these high temperatures occur at distances less than one AU -- often much less.  
Figure \ref{fig:p_tmax}b shows that the maximum temperatures, and the proportions of the temperatures higher than specific values generally decrease with radius.

There is a peak in Figure \ref{fig:p_tmax}b for small radii ($\textless 0.5$ AU) which means there is a small subset of systems with high temperatures there.  The main reason for this effect is that with high $T_{\rm C}$ and low $\omega_{\rm C}$, the infall rate from the MCC to the disc is high.  With low viscosity in the disc, a lot of material accumulates in the inner region of the disc.  The resulting high surface density raises the opacity and absorption of energy from the central star, resulting in high temperatures (See also Figure \ref{fig:T-t} and \ref{fig:tmax_diff}).

\section{Discussion}
\label{sec:discussion}

\subsection{Influence on composition of chondrites and terrestrial planets}

Observations show that the compositions of CM, CO, and CV chondrites and terrestrial planets are substantially depleted in volatile and somewhat enriched in refractory elements relative to the composition of the Sun \citep[][see also Figure 1 in \citet{Li:2020}]{Asplund:2005,Asplund:2009}.  Partial condensation of the elements in a cooling disc from very high temperature are thought to be responsible for this depletion pattern \citep{Grossman:1972,Cassen:1996,Li:2020}.

These studies assume that the initial temperature in the disc is higher than the condensation temperature of most refractory elements -- resulting in nearly complete vaporization of presolar MCC dust.  The chemistry then evolves through a single condensation sequence from initially gaseous material.  However, our calculations indicate that the maximum temperatures in most discs are lower than the condensation temperature of most refractory elements.  This means that, the refractory and some moderately volatile elements in most discs will remain in the primordial MCC dust. 
These elements would not fractionate, and the relative elemental abundances of the dust would match the stellar composition.

However, the prediction that moderately volatile elements are not fractionated from refractory elements in the condensed phases (planetesimals) is not consistent with what we observe in the solar system.  In order to reproduce the observations of solar system objects, there must be some process or event that alters the condensation sequence.  We, therefore, examine these discs in more detail to see what observations can be explained with our models, and what observations require some new process.

In our calculations, we treat the fact that the dust may not condense at the same time as follows.  The chemical evolution of the disc uses the method described in \citet{Li:2020}.
In all our models, the mid-plane temperatures are not hot enough to completely evaporate all of the dust, so that both the dust and gas exist at $t=0$.  We assume that the combination of the condensed phases and the gas immediately reaches chemical equilibrium, but the dust isolation (decoupling) from the gas phase takes time.  This decoupling of the dust is modelled with a decoupling timescale (see \citet{Li:2020}).  With this approach, the assumption is that the infall dust grains from the MCC are small enough to quickly reach chemical equilibrium under the $P-T$ conditions of the disc.

With the approach outlined above, only a small subset of our initial MCC conditions are capable of reproducing the depletion pattern of the chondrites and terrestrial planets (and then, the results are still a poor fit to the observed element patterns of CM, CV, and CO chondrites).  These results imply that the initial conditions of the solar system are either rare, or we need other energy sources to heat the disc to a very high temperature to reset the condensation sequence.

Moreover, the maximum temperatures in our simulations occur at radii less than 1 AU.  Not only is it rare to reach high enough temperatures to reset the condensation sequence, but the composition of planets beyond 1 AU can only be affected by that condensation if sufficient materials are transported outwards from the inner $\sim 0.5$AU.  For many discs (including our fiducial model) some material is indeed transported outward from the interior to 1 AU (see Figure \ref{fig:discevo}).  But, that material would mix with previously condensed and decoupled materials in the outer regions -- somewhat diluting the signatures of the chemical evolution of the inner region.

\subsection{Influence on the formation of compounds}

\begin{figure}
	\includegraphics[width=0.5\columnwidth]{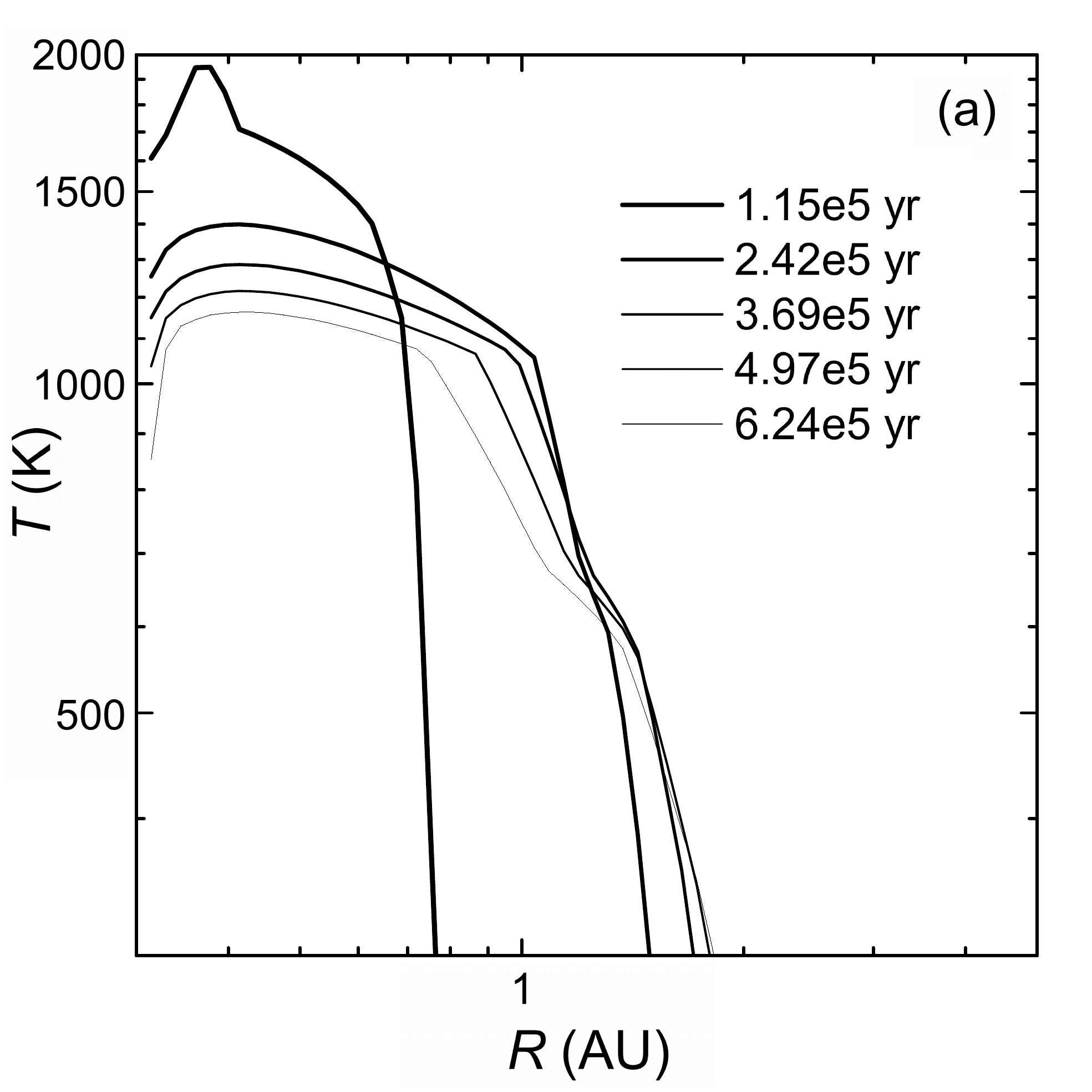}
	\includegraphics[width=0.5\columnwidth]{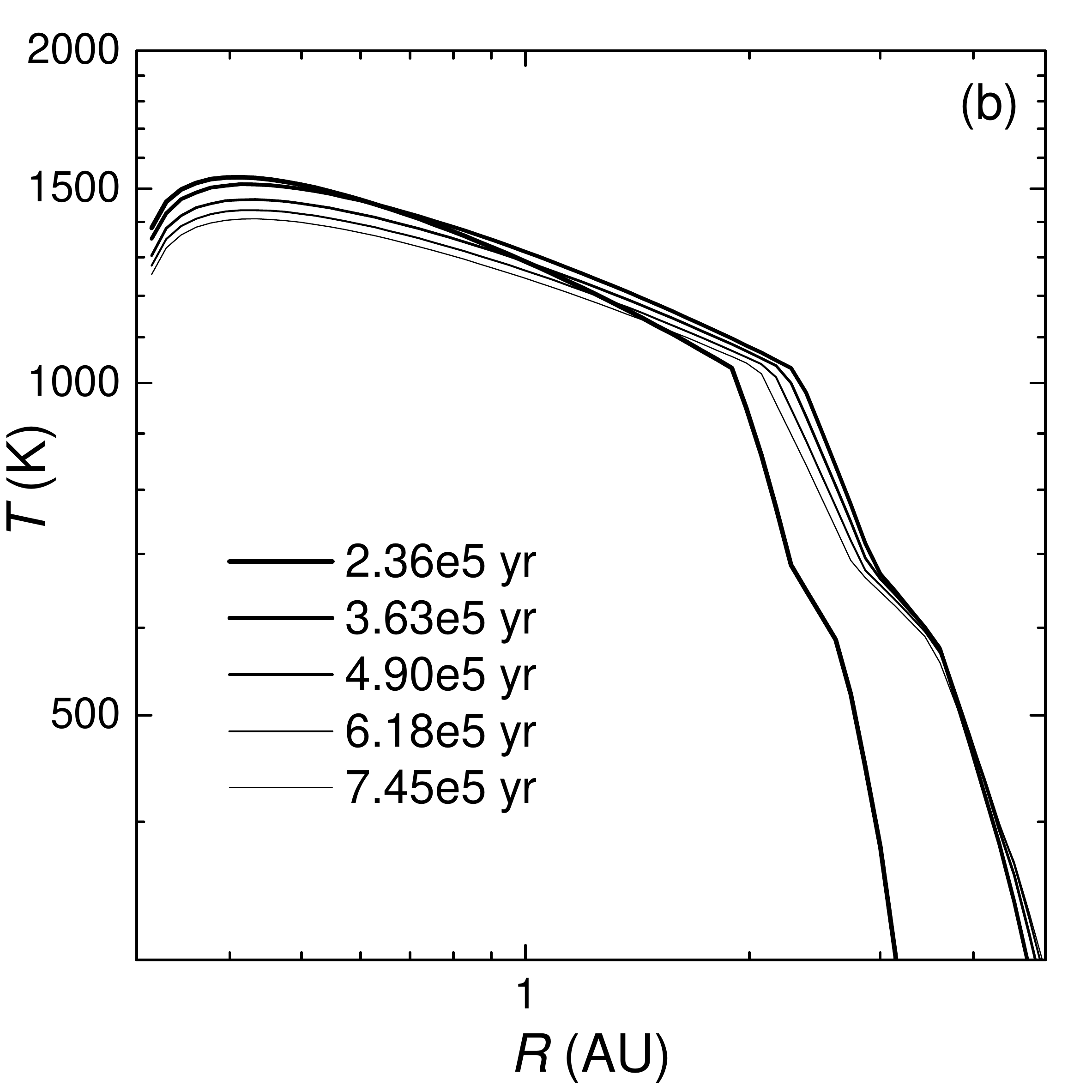}
	\includegraphics[width=0.5\columnwidth]{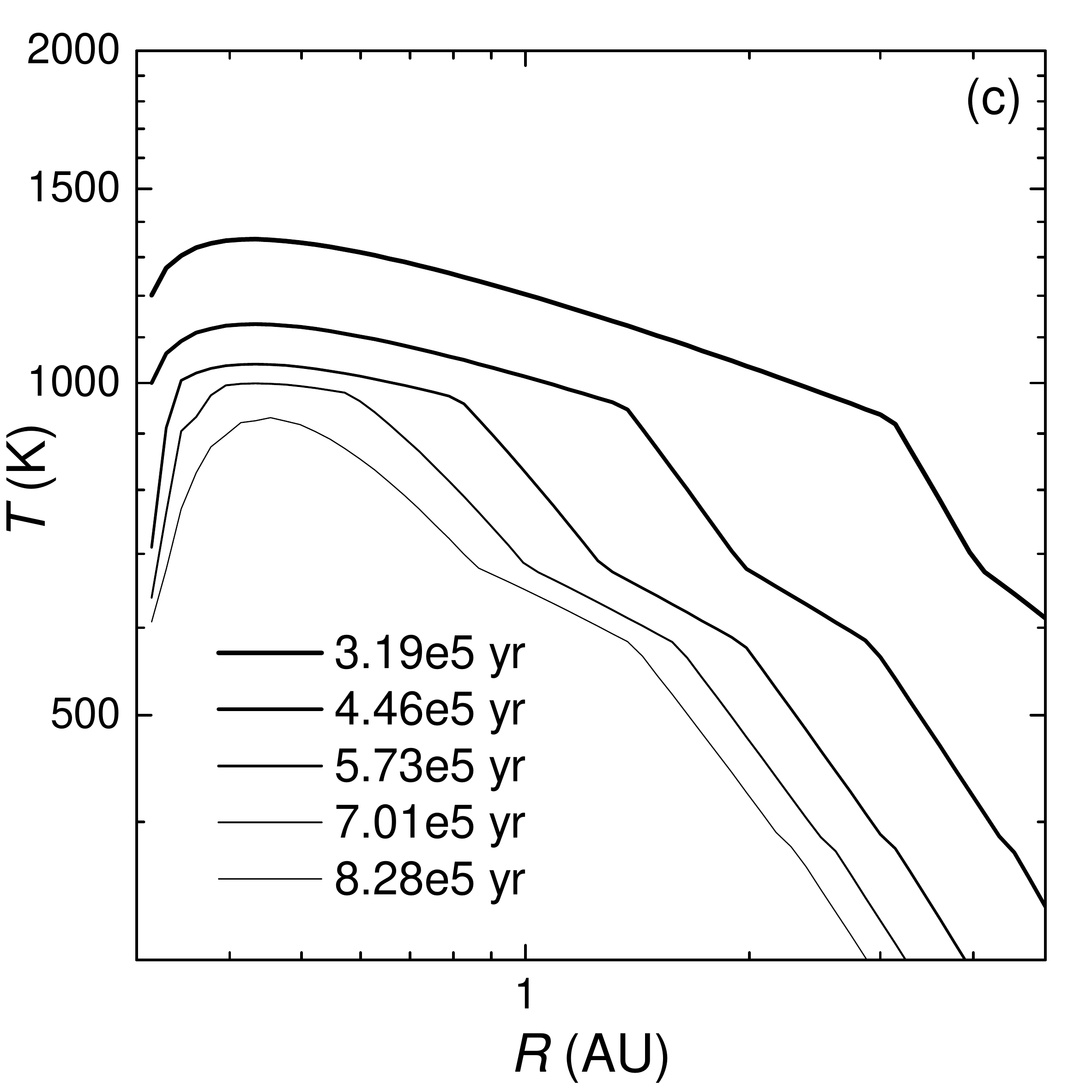}
	\includegraphics[width=0.5\columnwidth]{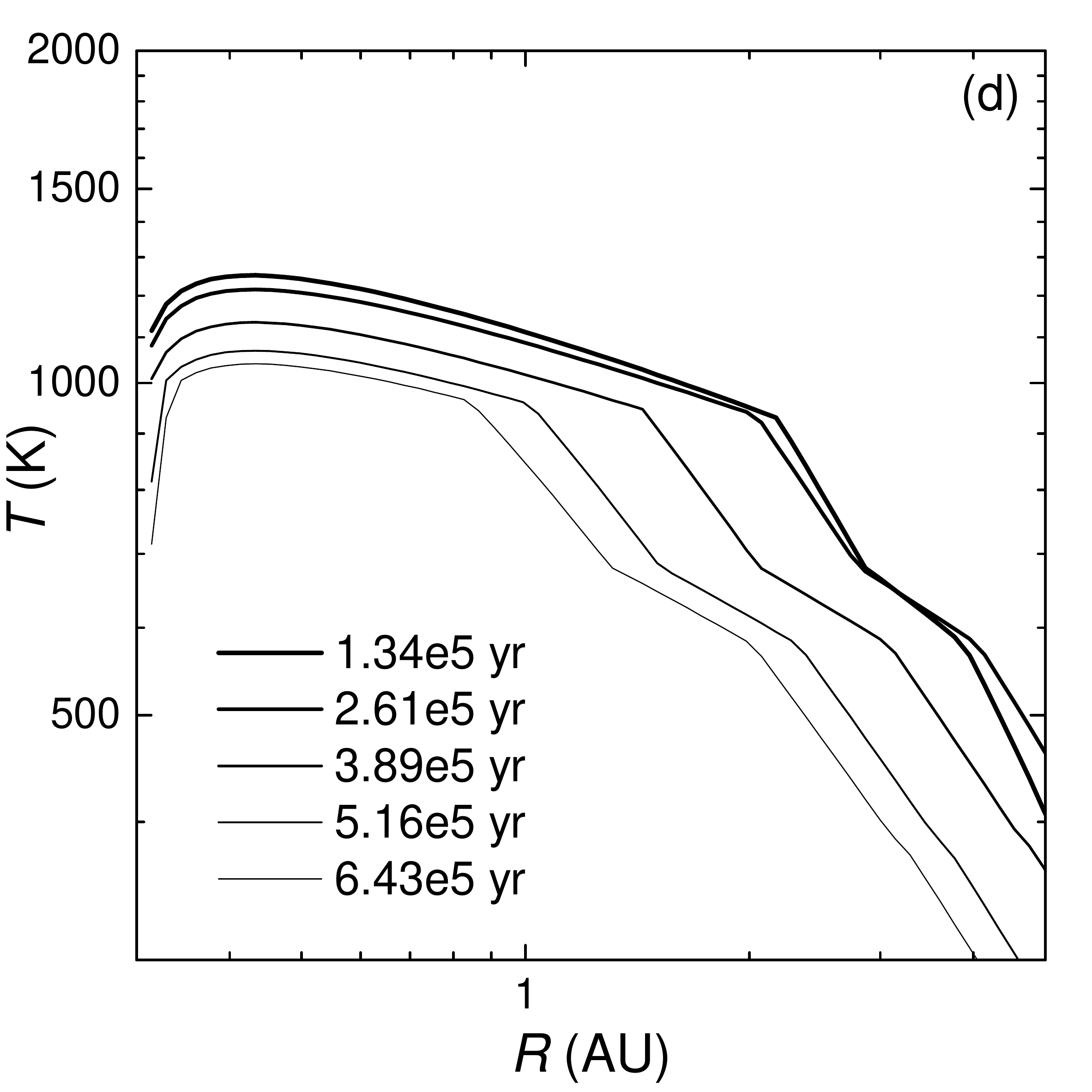}
    \caption{Temperature evolution for different initial conditions. (a)  $T_{\rm C}=31$ K, $\omega_{\rm C}=1\times 10^{-14} \rm\ s^{-1}$, and $\alpha=10^{-5}$,
    (b) $T_{\rm C}=15$ K, $\omega_{\rm C}=1\times 10^{-14} \rm\ s^{-1}$, and $\alpha=10^{-5}$,
    (c) $T_{\rm C}=15$ K, $\omega_{\rm C}=1\times 10^{-14} \rm\ s^{-1}$, and $\alpha=10^{-3}$, and
    (d) $T_{\rm C}=15$ K, $\omega_{\rm C}=3\times 10^{-14} \rm\ s^{-1}$, and $\alpha=10^{-3}$.
    For all the cases,     
    $M_{\rm C}=1\ M_\odot$.
    The time for the first line here are set to be the time when the temperature gets its maximum value. The time interval is 20\% of the collapse time of MCC for $M_{\rm C}=1\ M_\odot$,  $T_{\rm C}=10$ K, and $\omega_{\rm C}=1\times 10^{-14} \rm\ s^{-1}$. 
    }
    \label{fig:tmax_diff}
\end{figure}

\begin{figure}
\centering 
	\includegraphics[width=0.8\columnwidth]{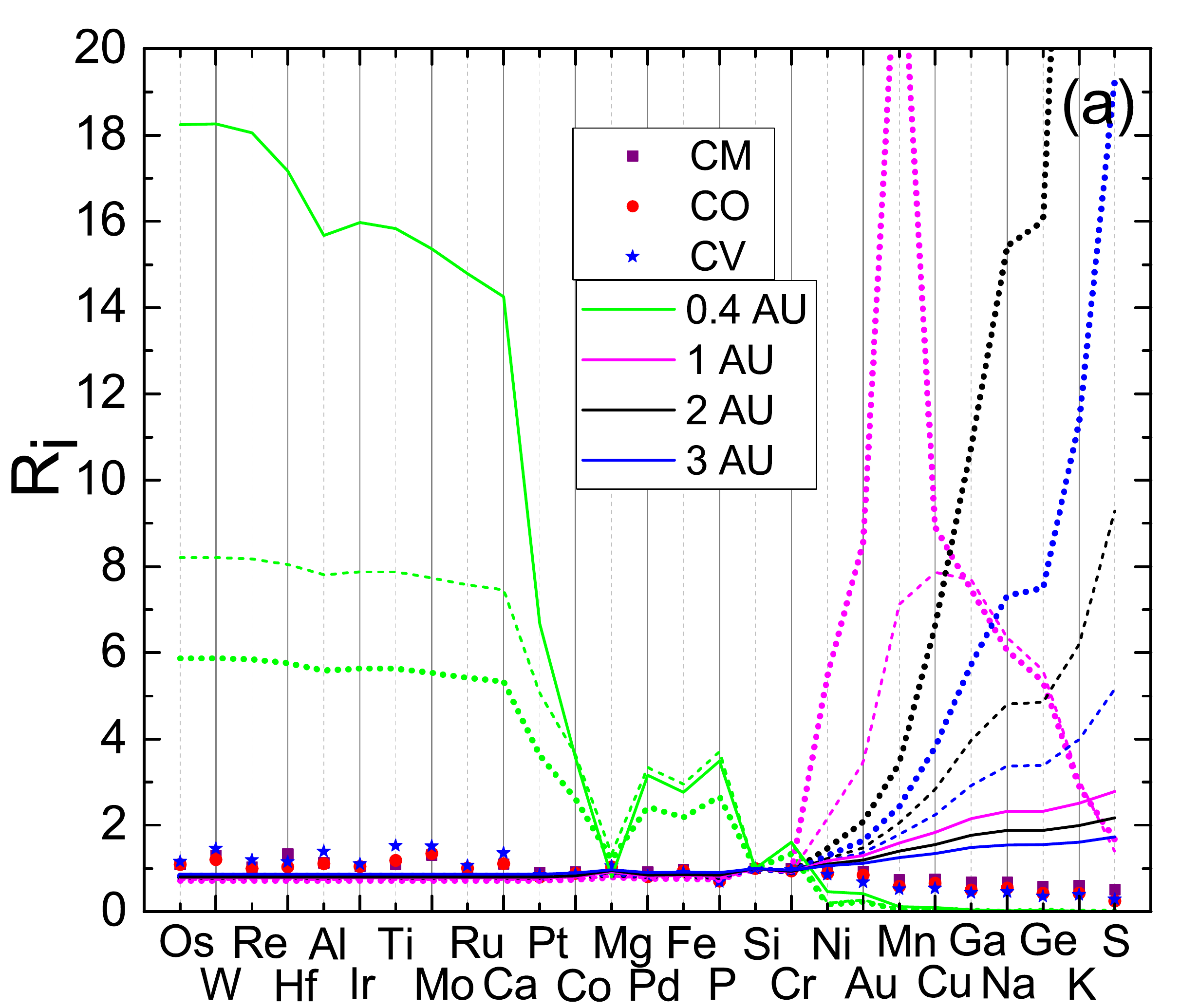}
	\includegraphics[width=0.8\columnwidth]{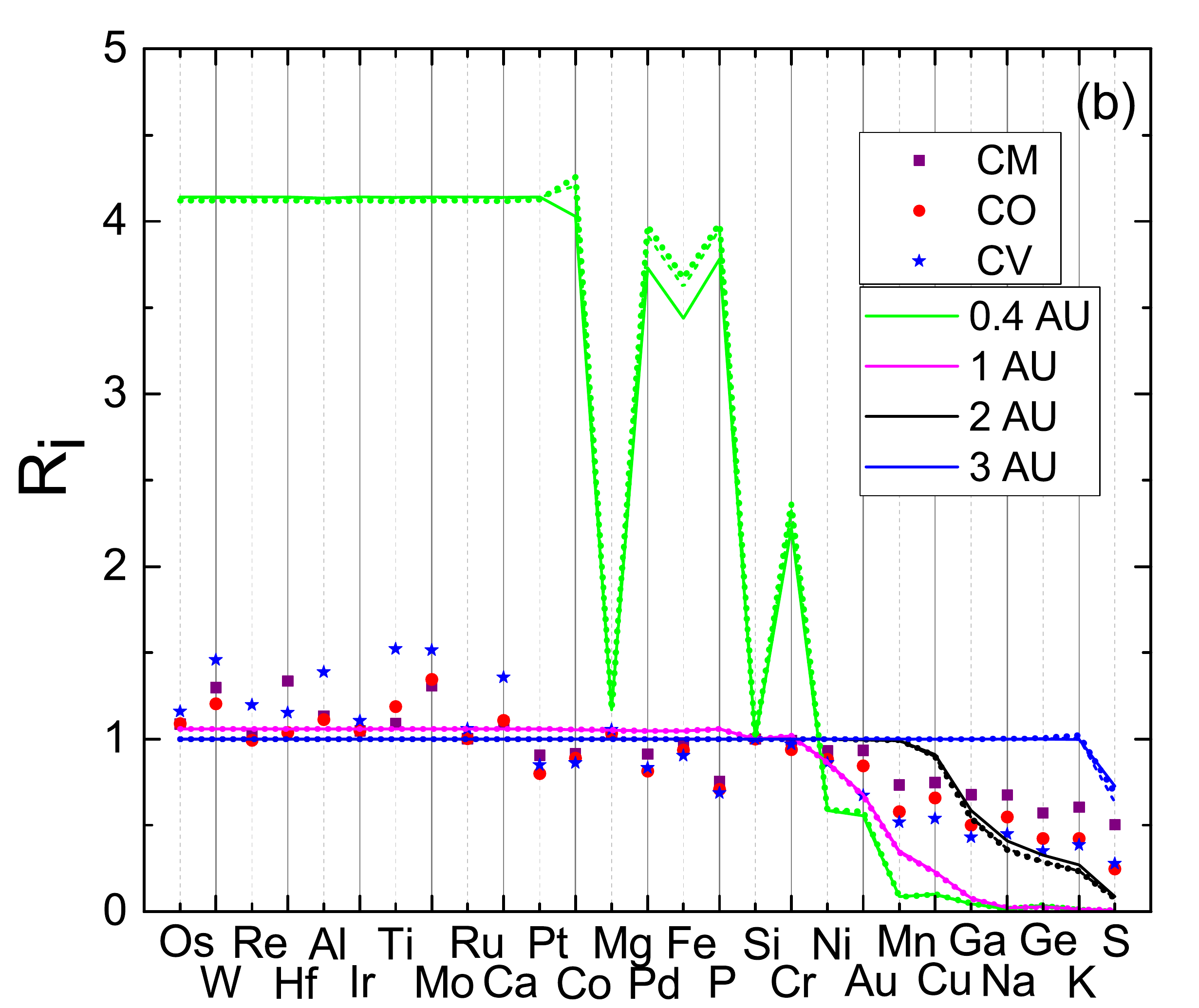}
	\includegraphics[width=0.8\columnwidth]{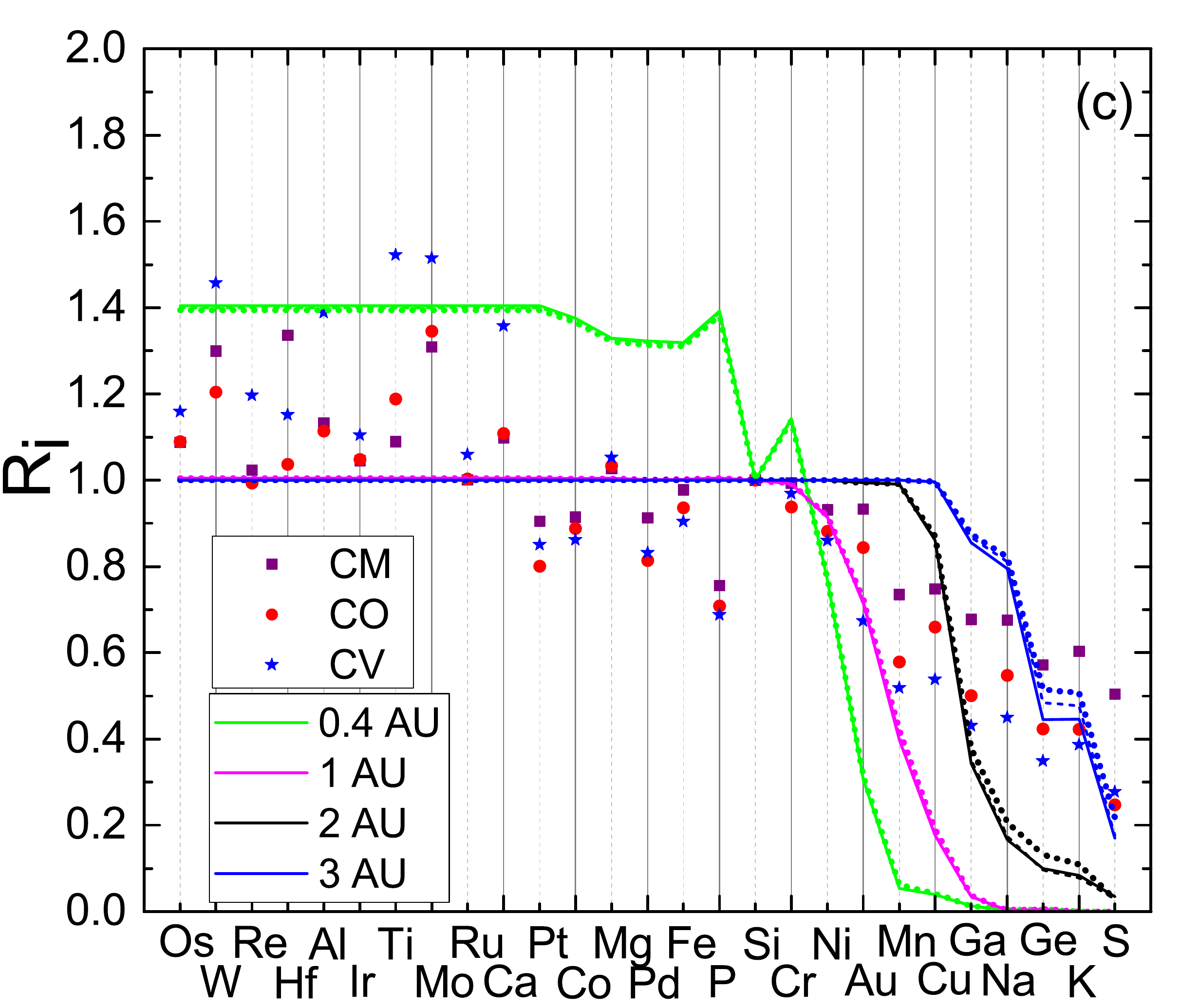}
	\includegraphics[width=0.8\columnwidth]{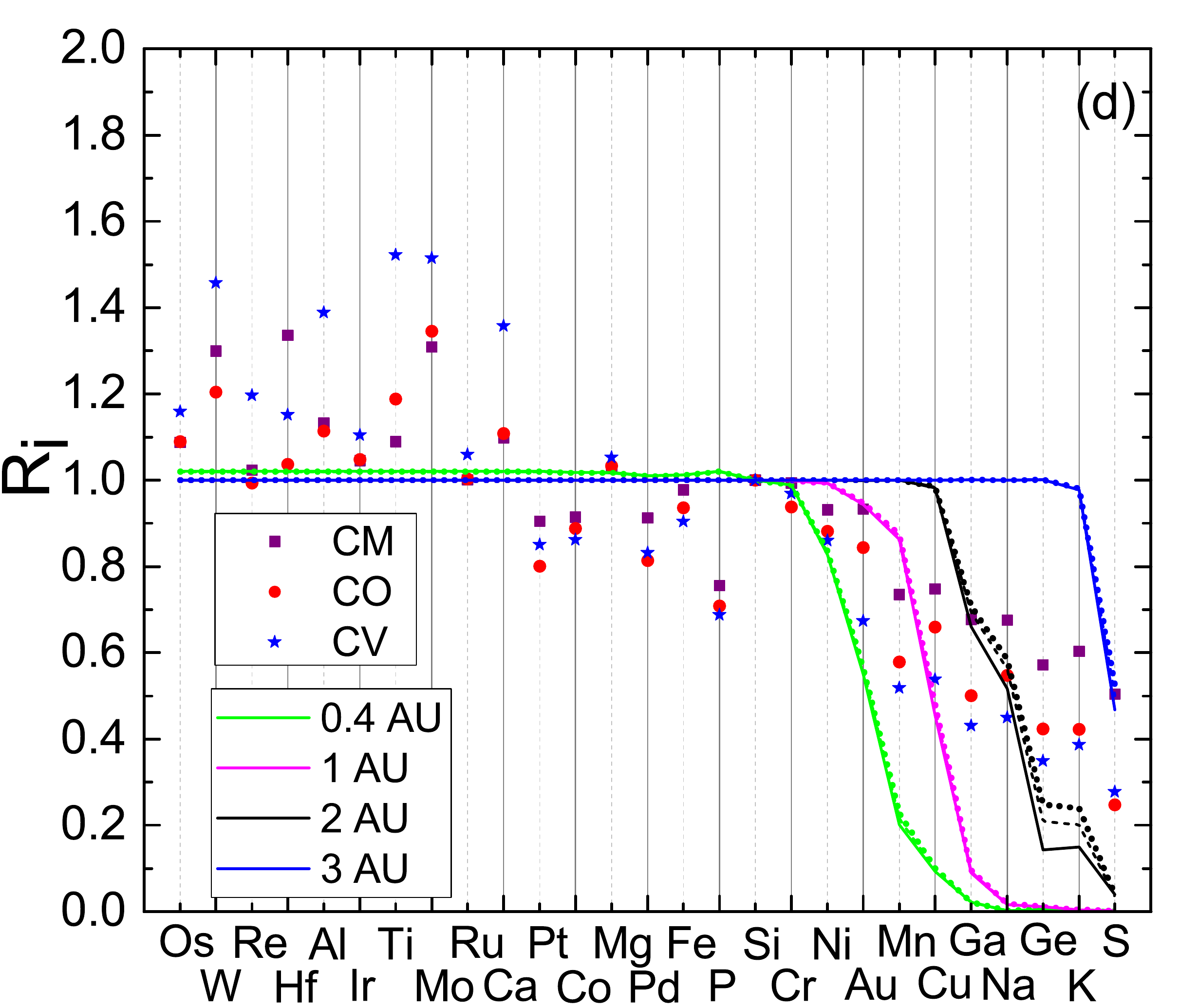}
    \caption{Elemental abundances at different radii for three moments in time, from the end of the collapse to 5$t_{\rm dec}$ from then. Initial conditions are the same as in Figure \ref{fig:tmax_diff}.
    }
    \label{fig:Riele}
\end{figure}

\begin{figure}
\centering 
	\includegraphics[width=0.85\columnwidth]{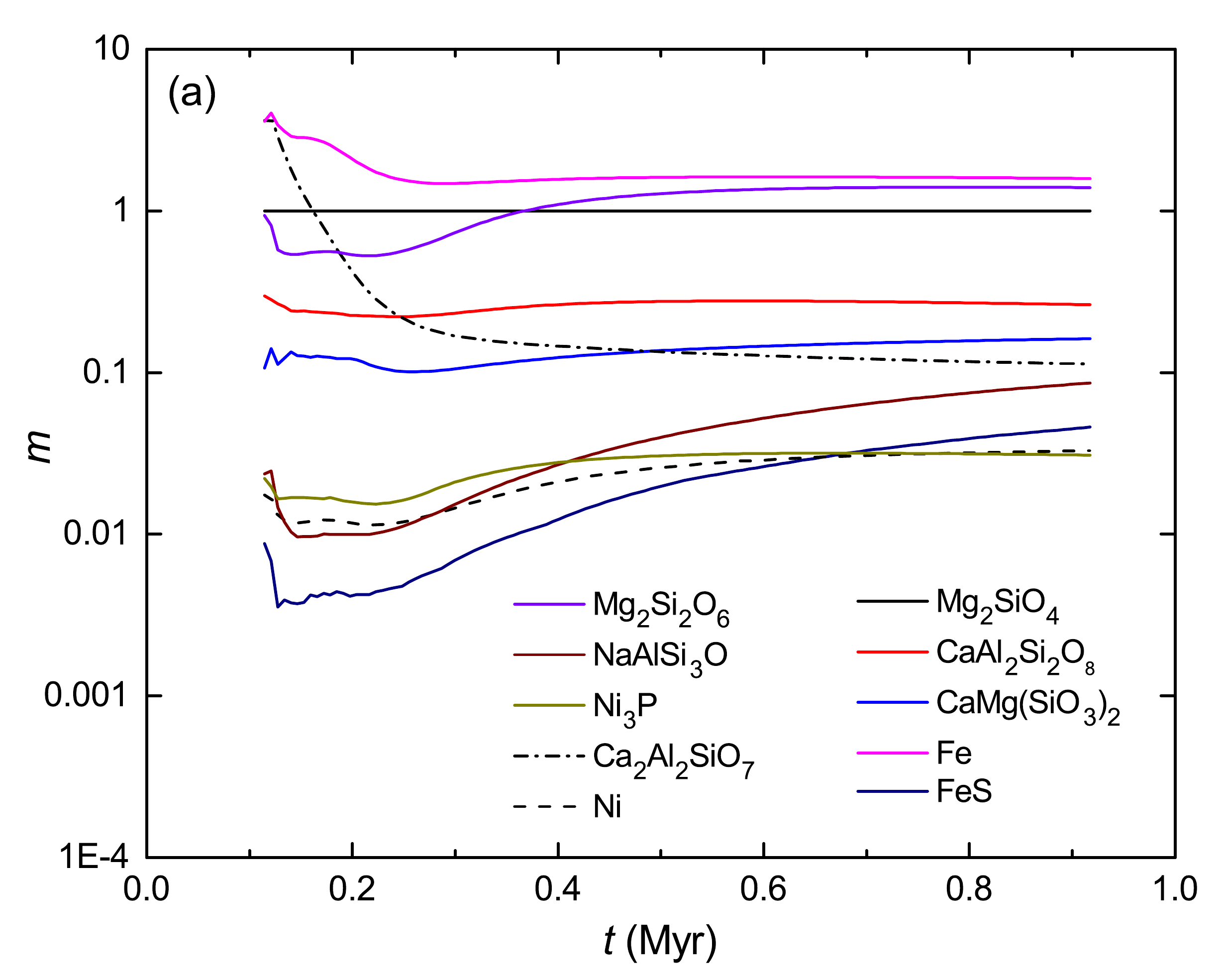}
	\includegraphics[width=0.85\columnwidth]{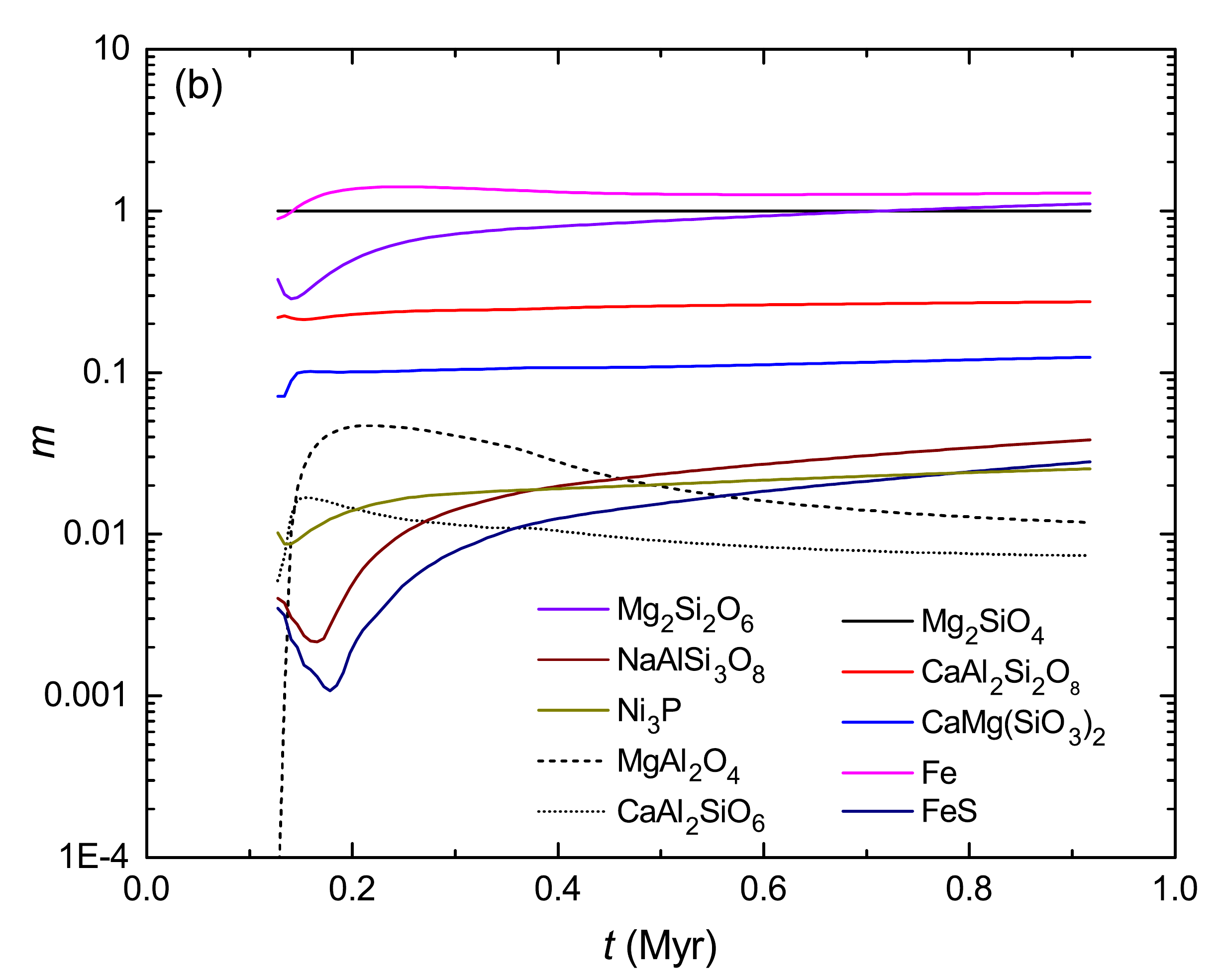}
	\includegraphics[width=0.85\columnwidth]{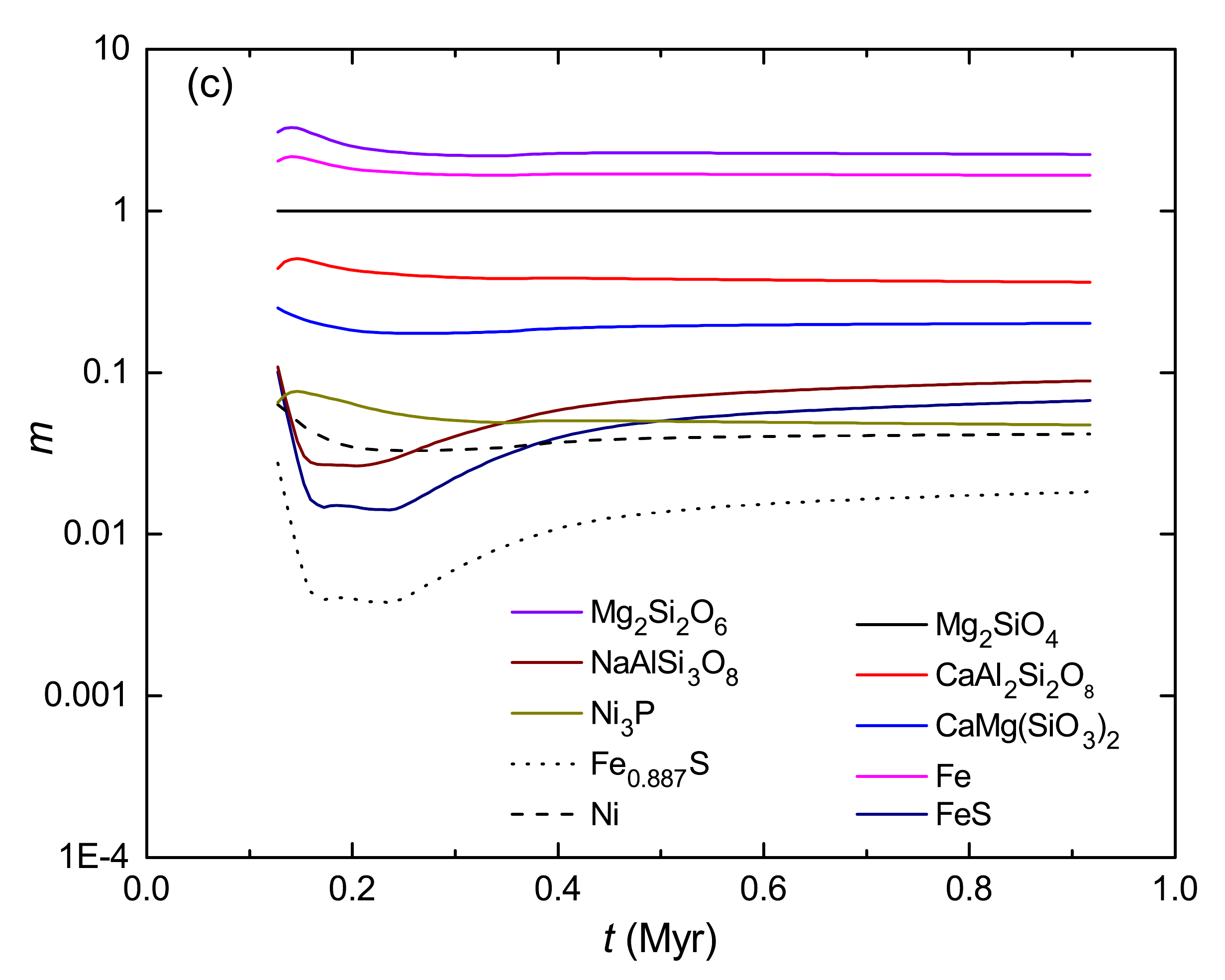}
	\includegraphics[width=0.85\columnwidth]{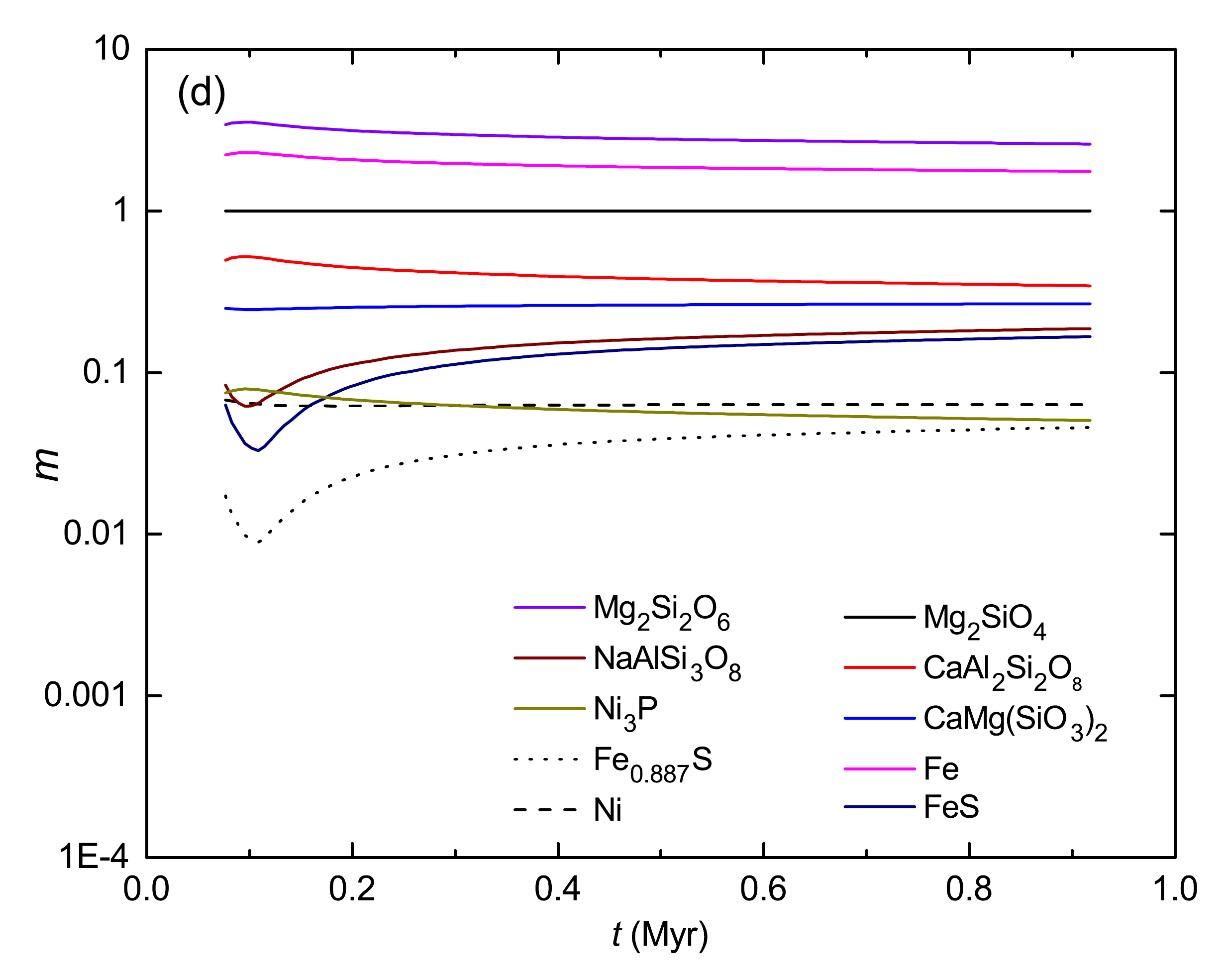}
    \caption{Evolution of the 10 most abundant condensed species, normalized to Mg$_2$SiO$_4$. The initial conditions are the same as in Figure \ref{fig:tmax_diff}. Solid lines represent species that are common to all panels while broken lines are unique to a specific panel.
    }
    \label{fig:species_diff_ratio}
\end{figure}

\begin{figure}
\centering 
	\includegraphics[width=0.85\columnwidth]{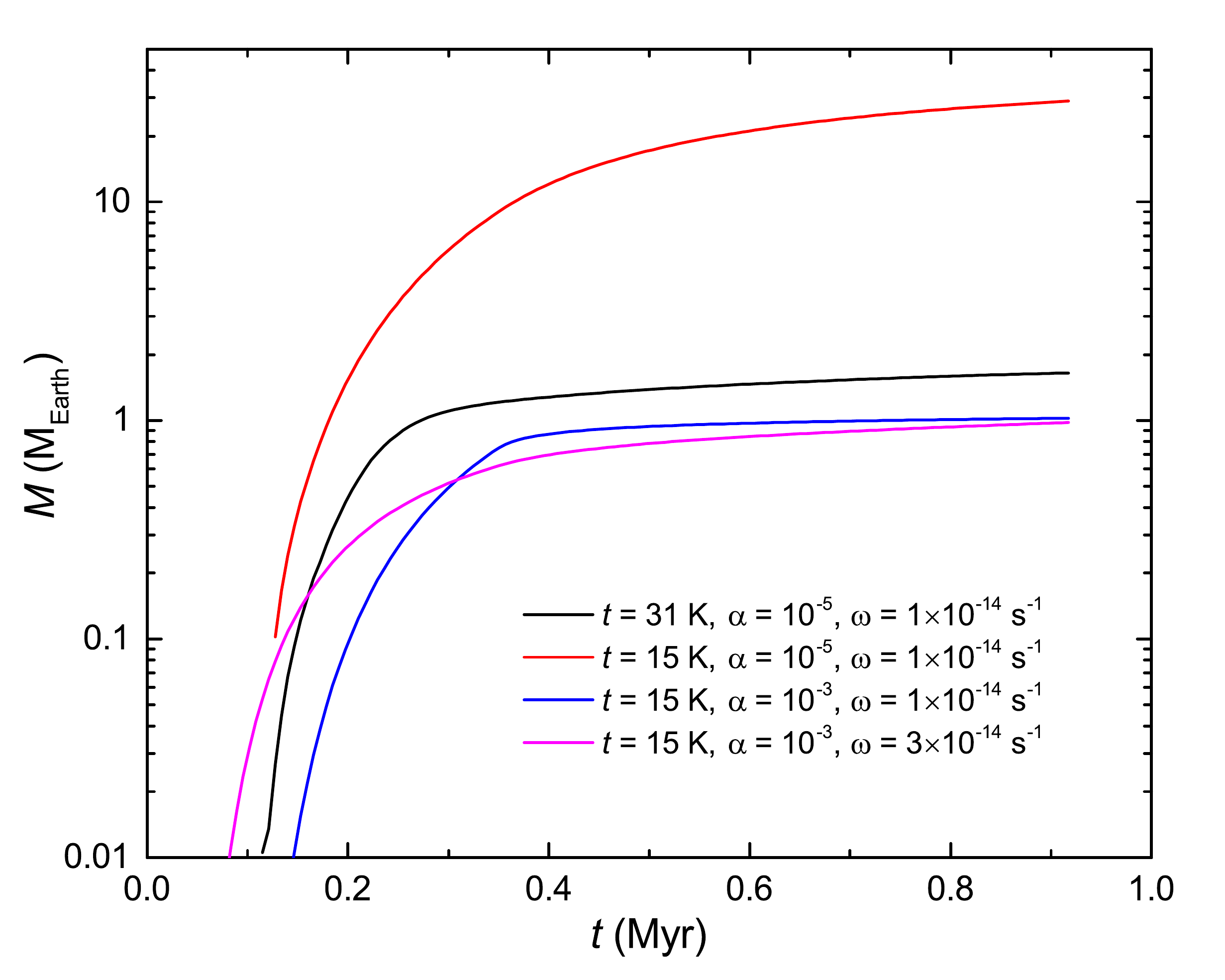}
    \caption{Mass evolution of Mg$_2$SiO$_4$ for the four initial conditions used in the other figures.
    }
    \label{fig:mt}
\end{figure}

\begin{figure}
	\includegraphics[width=\columnwidth]{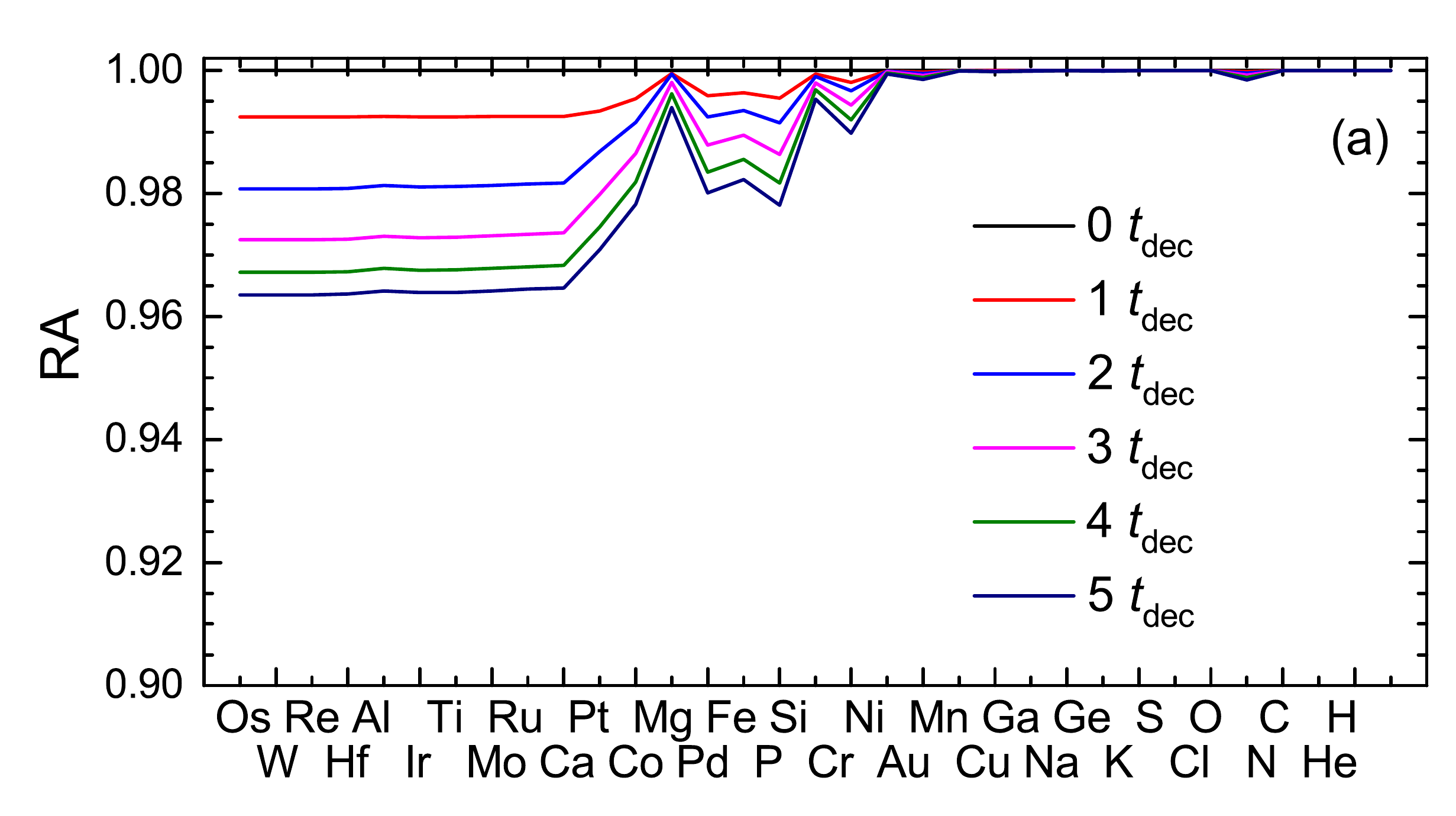}
	\includegraphics[width=1.0\columnwidth]{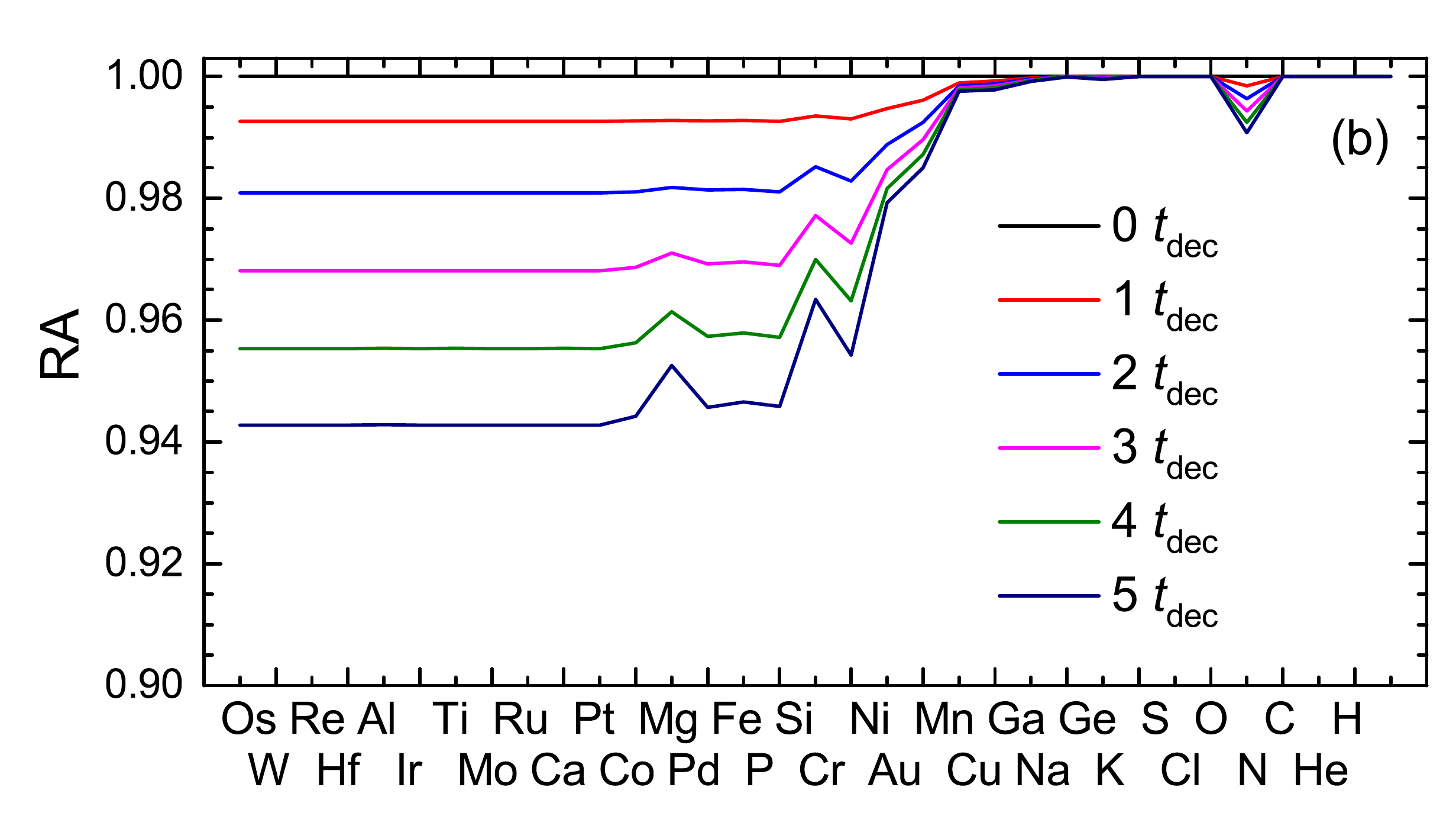}
	\includegraphics[width=1.0\columnwidth]{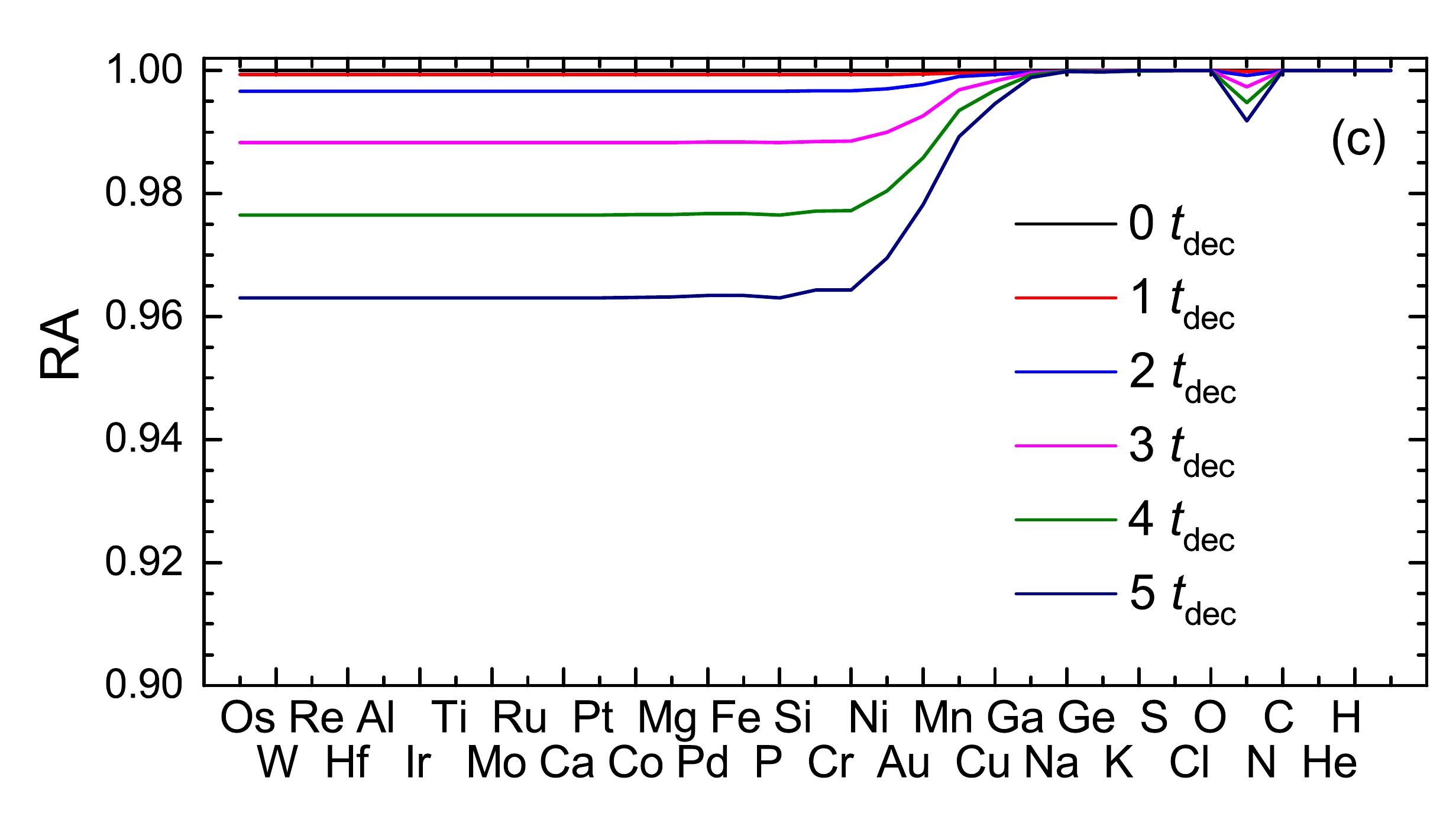}
	\includegraphics[width=1.0\columnwidth]{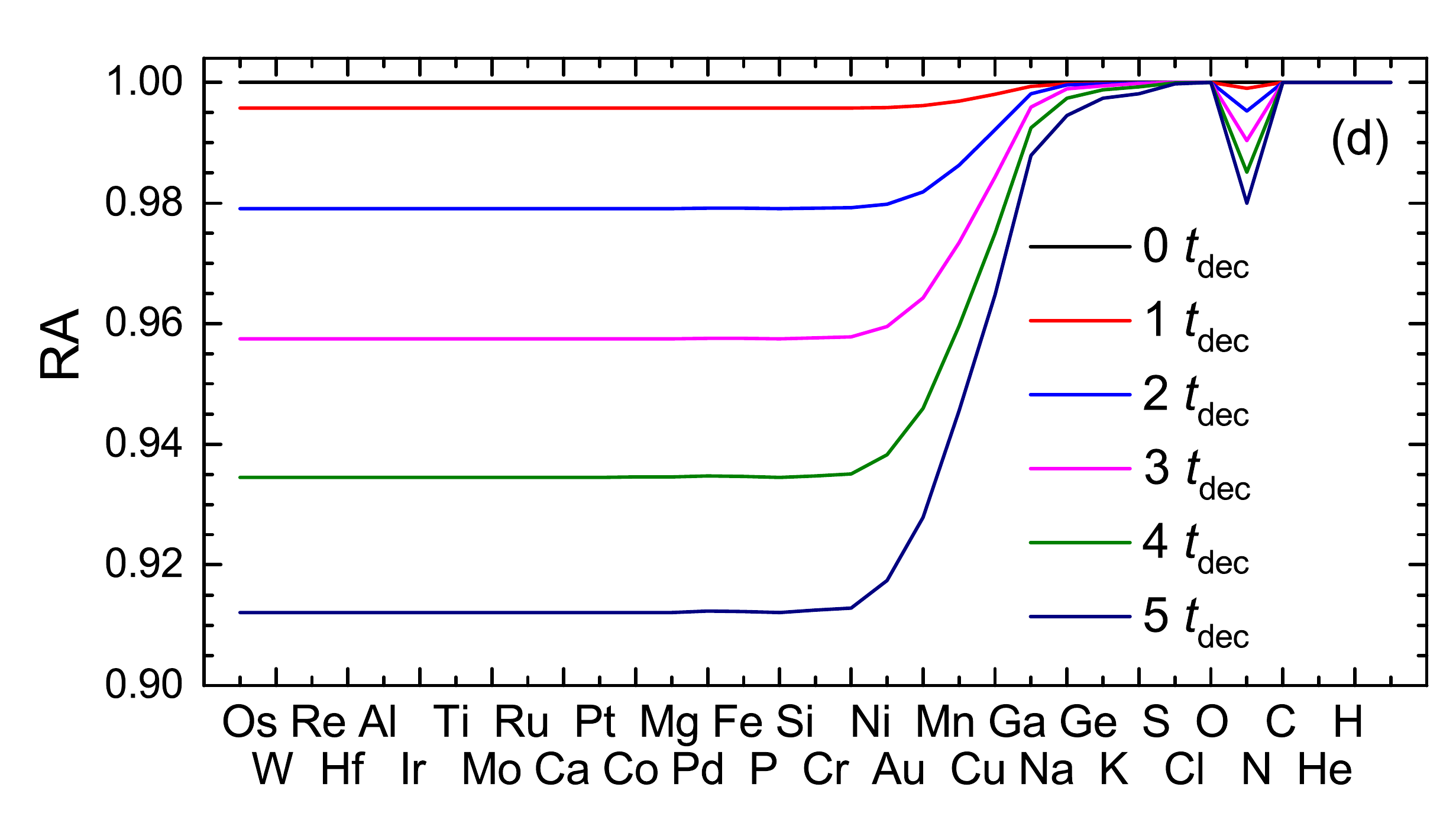}
    \caption{Relative element abundances (RA, using solar abundance as the reference) in the central star as a function of time. 
    The initial conditions are the same as in Figure \ref{fig:tmax_diff}.
    (a)  $T_{\rm C}=31$ K, $\omega_{\rm C}=1\times 10^{-14} \rm\ s^{-1}$, and $\alpha=10^{-5}$,
    (b) $T_{\rm C}=15$ K, $\omega_{\rm C}=1\times 10^{-14} \rm\ s^{-1}$, and $\alpha=10^{-5}$,
    (c) $T_{\rm C}=15$ K, $\omega_{\rm C}=1\times 10^{-14} \rm\ s^{-1}$, and $\alpha=10^{-3}$, and
    (d) $T_{\rm C}=15$ K, $\omega_{\rm C}=3\times 10^{-14} \rm\ s^{-1}$, and $\alpha=10^{-3}$.
    For all the cases,     
    $M_{\rm C}=1\ M_\odot$. The initial time is the beginning of the evolution of the discs.
    }
    \label{fig:elet}
\end{figure}

To see the effects that these maximum temperatures have on the composition of the condensed material (i.e., planetesimals) we calculate their composition for discs that form from our MCCs assuming initial solar abundances.  We calculate the chemical equilibrium of the 33 elements \citep{Petaev:2009,Li:2020} at each radius 
and each time in the evolving disc. We use a decoupling time-scale of 1.5$\times 10^{4}$ yr, which is consistent with the standard value used by \citet{Li:2020}.  While the decoupling timescale is two orders-of-magnitude shorter than the disc evolution timescale, the compositions of the decoupled materials are significantly affected by the radial transport \citep{Li:2020}.

The temperature evolution of four discs with different initial conditions are shown in Figure \ref{fig:tmax_diff}.  The first lines are set to the times when the temperatures reach their maximum values.  The time intervals for different lines are 20\% of the typical collapse timescale.  The temperatures in Figure \ref{fig:tmax_diff} reach their maximum values at distances between 0.38 and 0.43 AU.

Figure \ref{fig:Riele} shows the relative elemental abundances of decoupled elements normalized to the Solar abundance and Si as a function of 50\% condensation temperature at different radii.  Generally, the relative abundances of refractory elements tend to be high in discs with a high maximum temperature (Figure \ref{fig:Riele}) and their abundances decrease with time as more volatile elements condense and decouple as the temperature decreases \citep{Li:2020}.  The relative abundance of elements with 50\% condensation temperatures higher than the maximum temperature at this radius are equal because they have never been vaporized (they are not fractionated from each other).  There are some non-monotonic changes for temperatures between 1200 to 1400 K in panels (a) to (c) (around the element Mg).  These arise because the elements with lower 50\% condensation temperatures may have higher relative abundances at certain disc midplane temperatures (See Figure 4 in \citet{Li:2020}). The abundance beyond 1 AU rarely strays from unity, even in the hottest disc models (except panel a). This occurs because, beyond the 1 AU, the disc temperature never reaches 1500 K.

Compared with previous results that use simple, analytic disc models, these more realistic disc models cannot reproduce the observed volatile element depletion patterns in CM, CO, and CV (Figure \ref{fig:Riele}).  To reproduce the observations, either additional heating events are needed or the volatile element depletion patterns in these chondrites were produced outside the protoplanetary disc, and they are simply a record of whatever falls into the protoplanetary disc.

Figure \ref{fig:species_diff_ratio} shows how the amounts of the 10 most abundant (by mass) condensed species change with time relative to Mg$_2$SiO$_4$ for discs shown in Figure \ref{fig:tmax_diff}
while the mass evolution of Mg$_2$SiO$_4$ is shown in Figure \ref{fig:mt}.  Most of these species are the same across all panels, but there are some differences.  For example, panel (a) includes Ca$_2$Al$_2$SiO$_7$, which is stable at the 1363 - 1505 K temperature range, while the other panels do not. The relative amounts of the species also differ from disc to disc, especially in (a) and (b).
We also see that the relative amount of each species tends to be stable for each disc -- which may help constrain the thermal history of the discs.

As the amount of condensed materials change for different MCC properties, the relative abundance of the final central star should also be different.  Figure \ref{fig:elet} shows the relative chemical abundance of the central star as a function of element and time if we assume that all decoupled materials stay in the region where they condense.  Initially, all of them have solar abundance.  Over time, more elements condense and the relative abundances of the decoupled material decreases.  The abundance for the refractory elements are lower than the volatile elements, which is consistent with the fact that more refractory elements decouple from the disc.  This indicates that the chemical abundances in stars will differ somewhat from the abundances of the MCC from which they form as material becomes trapped in the planets. 
A consequence of this result is that, when we use the stellar composition as a proxy for the composition of planets, we may need to use a composition somewhat enriched in refractory elements.

\section{Conclusions}\label{sec:conclusion}

Our results show that the properties of the initial MCCs have a significant effect on the maximum temperatures that are reached within the resulting discs.  The disc temperatures, in turn, affect the composition of chondrites and planets that form.  These simulations lead to the following conclusions:

1. The temperature in the inner region ($\textless10$ AU) of the disc increases first and then decreases with time.  It reaches its maximum value around the end of the MCC collapse (several times $10^5$ years). 

2. The maximum temperature of the disc increases with the initial temperature of the MCC and decreases with its angular velocity.  The maximum temperature in the disc also increases if the viscosity in the disc decreases since the additional material in the inner region traps more heat in the disc.

3. The resulting composition of the central stars are similar to the initial composition of the MCC.  However, the central stars might be slightly depleted in some of the most refractory elements -- by up to 10\% compared to the initial composition.  Consequently, stellar composition may or may not be a good approximation to the initial composition of the MCC from which it formed, depending upon the situation.

4. 90\% of the simulations that use the observed properties of MCCs predict peak temperatures between 935 and 1635 K, with a median value of 1250 K.  Less than 1\% of the discs from our simulations reach temperatures higher than 2000 K and are capable of vaporizing all elements.  Even in these simulations, those temperatures only arise in the innermost portion of the disc.
Note that these peak temperatures depend upon our chosen distribution of $\alpha$ values.

5. Most discs reach peak temperatures that are lower than the 50\% condensation temperature of Mg.  To match the depletion patterns of CM, CO, and CV chondrites and terrestrial planets, one needs either rare initial conditions of the proto-solar MCC, or other energy sources to heat the disc to a very high temperature in order to
reprocess the moderately volatile to refractory elements.

\section*{Acknowledgements}

JHS, ML, and SH are supported by the NASA grant NNX16AK08G and NSF grant AST-1910955.  ZZ acknowledges support from the National Science Foundation under CAREER Grant Number AST-1753168 and Sloan Research Fellowship.

\section*{DATA AVAILABILITY}

The data underlying this article are available in the article.

%%%%%%%%%%%%%%%%%%%%%%%%%%%%%%%%%%%%%%%%%%%%%%%%%%



\bibliographystyle{mnras}
\bibliography{tem} % if your bibtex file is called example.bib

%\begin{thebibliography}{99}
%\end{thebibliography}

%%%%%%%%%%%%%%%%%%%%%%%%%%%%%%%%%%%%%%%%%%%%%%%%%%

%%%%%%%%%%%%%%%%% APPENDICES %%%%%%%%%%%%%%%%%%%%%

%\appendix


%If you want to present additional material which would interrupt the flow of the main paper, it can be placed in an Appendix which appears after the list of references.

%%%%%%%%%%%%%%%%%%%%%%%%%%%%%%%%%%%%%%%%%%%%%%%%%%

% Don't change these lines
\bsp	% typesetting comment
\label{lastpage}
\end{document}